\documentclass[13pt,preprint]{aastex}
\slugcomment{{\sc Accepted to ApJ:} June 6, 2014}
\usepackage{graphicx}
\usepackage{rotating}
\usepackage{booktabs}
\usepackage[verbose]{placeins}
\usepackage[ngerman]{babel}
\usepackage{blindtext}
\usepackage[caption=false]{subfig}
\usepackage{amsmath}

\usepackage{array}
\newcolumntype{+}{>{\global\let\currentrowstyle\relax}}
\newcolumntype{^}{>{\currentrowstyle}}

\begin{document}

\title{A Search for Fast Radio Bursts Associated with Gamma-Ray Bursts}

\author{ Divya $\rm Palaniswamy^{1}$,  Randall B. $\rm Wayth^{1}$,  Cathryn M. $\rm Trott^{1,2}$, Jamie N. $\rm McCallum^{3}$,\\ Steven J. $\rm Tingay^{1}$ , Cormac  $\rm Reynolds^{1}$}
\affil{\textit{1. International Centre for Radio Astronomy Research, Curtin University, Bentley WA, 6102}}
\affil{\textit{2. ARC Centre of Excellence for All-Sky Astrophysics (CAASTRO)}}
\affil{\textit{3. University of Tasmania, Hobart, Tasmania, 7001}}

\email{divya.palaniswamy@postgrad.curtin.edu.au}

\renewcommand{\abstractname}{Abstract}

\begin{abstract}

The detection of six Fast Radio Bursts (FRBs) has recently been reported. FRBs are short duration ($\sim$ 1 ms), highly dispersed radio pulses from astronomical sources. The physical interpretation for the FRBs remains unclear but is thought to involve highly compact objects at cosmological distance. It has been suggested that a  fraction of FRBs could be physically associated with gamma-ray bursts (GRBs). Recent radio observations of GRBs have reported the detection of two highly dispersed short duration radio pulses using a 12 m radio telescope at 1.4 GHz. Motivated by this result, we have performed a systematic and sensitive search for FRBs associated with GRBs. We have observed five GRBs at 2.3 GHz using a 26 m radio telescope located at the Mount Pleasant Radio Observatory, Hobart. The radio telescope was automated to rapidly respond to Gamma-ray Coordination Network notifications from the Swift satellite and slew to the GRB position within $\sim$ 140 s. The data were searched for pulses up to 5000 pc $\rm cm^{-3}$ in dispersion measure and pulse widths ranging from 640 $\rm \mu$s to 25.60 ms. We did not detect any events $\rm \geq 6 \sigma$. An in-depth statistical analysis of our data shows that events detected above $\rm 5 \sigma$ are consistent with thermal noise fluctuations only. A joint analysis of our data with previous experiments shows that previously claimed detections of FRBs from GRBs are unlikely to be astrophysical. Our results are in line with the lack of consistency noted between the recently presented FRB event rates and GRB event rates.

\end{abstract}
 
\section{Introduction}

Gamma-ray bursts (GRBs) are among the most powerful objects in the Universe. The energy output in the $\gamma$-rays is in the order of $10^{51}$ ergs (assuming isotropic emission) on second time-scales at cosmological distances \citep{Rees1994,Ghir2006}. GRBs are divided into two classes, short and long, which are usually defined by $\rm T_{90}$ \citep{Kouveliotou1993}, the time during which the cumulative counts in the detector increases from 5$\%$ to 95$\%$ above the background rate. There are two types of plausible central engine model proposed for GRBs: (1) stellar mass black holes, which accrete material from a remnant star with an extremely high accretion rate ($\sim$ (0.1-1) $\rm M_{\bigodot}$ $\rm s^{-1}$) \citep{Popham1999, Narayan2001, Lei2013}; (2)  rapidly rotating ($\sim$ ms duration), highly magnetised neutron stars (a proto-magnetar) \citep{Usov1992,Row2010,Fan2013}. Based on the latter model of the central engine, \citet{Usov2000} proposed that GRBs may be accompanied by very bright, short duration ($\leqslant$ ms) radio bursts, which could be observed at frequencies of tens of MHz. Detection of such radio emission from a GRB at a cosmological distance could provide a probe for understanding both the ionised intergalactic medium (IGM) and GRB physics \citep{Palmer1993, Ioka2003, Ghir2006}. However, \citet{JP2007} showed that observations of short duration radio pulses from GRBs at low radio frequencies ($\rm \sim 10-100$\,MHz) would be significantly limited by scattering effects because the radio emission passes through the dense stellar wind of the progenitor or  the immediate interstellar medium surrounding the source. \citet{JP2007} showed that these propagation effects alter the properties of radiation with brightness temperature $\gg \, 10^{10}$\,K, which may hinder the detectability of any radio pulse.

Recently, \citet{Zhang2014} proposed that GRBs (with a magnetar as the central engine) could produce short duration radio pulses when the supra-massive neutron stars undergo rapid spin down  (due to strong magnetic fields) within $\rm 10^{2}-10^{3}$ s of their birth, collapsing into  black holes as they lose centrifugal support. These short duration radio pulses are proposed to be detectable at radio frequencies ranging from $\rm 1.2 $ to $ \rm 1.5$ GHz, if they are not absorbed by the GRB blast wave in front of the emission region. \citet{Zhang2014} suggests that a small fraction of short duration radio pulses could be physically connected with GRBs.

In recent years, there have been number of detections of non-repeating short duration radio pulses. The dispersion measures (DM) of these pulses are so large that the progenitors are suggested to be extragalactic in origin. \citet{Lorimer2007} reported the discovery of the first such short duration radio pulse; they discovered a single pulse of duration $\rm \sim$ 5\,ms at a DM of 375 pc $\rm cm^{-3}$ from  archival data \citep{M2006} recorded using the 64 m Parkes radio telescope. \citet{Lorimer2007} inferred an event rate of 50 $\rm\, day^{-1} \, Gpc^{-3}$ (assuming an isotropic distribution of the sources in the sky) and proposed that the implied rates of occurrence were compatible with GRB rates. However, there are no recorded GRB events\footnote{In this case, a GRB event could have been simply missed, since no $\gamma-ray$ satellite was observing at the time and position of the event.} near the pulse location. Another event was discovered by \citet{Keane2011} from archival data \citep{M2001} with the same telescope. The pulse properties were similar to that reported by \citet{Lorimer2007}, with a short time scale ($\sim$ 7 ms) and high DM (745 pc $\rm cm^{-3}$). Recent analysis of this pulse has suggested it is Galactic in origin \citep{Bannister2014}

Some doubts on the extragalactic origin of these pulses have been put forward, based on the work of \citet{Burke-Spolaor2011}, with the detection of 16 similar highly dispersed pulses from Parkes archive data \citep{E2001,R2002,C2006,J2009}, all deemed as terrestrial in origin. Most recently, however, the evidence for an extragalactic origin has strengthened significantly, with the discovery of four new highly dispersed pulses by \citet{Thornton2013}. \citet{Thornton2013} coined the term Fast Radio Burst (FRB) to describe these bursts. The FRBs were detected with  DMs in the range 553 to 1103 pc $\rm cm^{-3}$. Based on their properties, \citet{Thornton2013} infers the FRB event rate ($\rm R_{FRB}$) to be $\rm 10^{-3} \, yr^{-1} \, galaxy^{-1}$ in the comoving sample volume that contains $\rm \sim \, 10^{9}$ late type galaxies, which is $\sim$ $10^{3}$ times larger than the GRB rates ($\rm R_{GRB}$ = $\rm 10^{-6} \, yr^{-1} \, galaxy^{-1}$). Using the inconsistency in the event rates and, in particular, with no known GRB events coincident with the FRBs, \citet{Thornton2013} rules out GRBs as  possible progenitors for most FRBs. 

All six FRBs mentioned above were detected by the 64 m Parkes radio telescope and its 13-beam receiver, creating concerns about the astrophysical origin of the pulses. Most recently, \citet{Spitler2014} reported a discovery of a seventh FRB in the 1.4 GHz Pulsar ALFA survey with the 300\,m Arecibo telescope, the first FRB discovered from a geographical location other than Parkes. This FRB was found at low galactic latitude ($\rm b = -0.2^{\circ}$) with a DM of $\rm 557.4 \, pc \, cm^{-3}$ and pulse width $\rm \sim 3 \, ms$. Despite the low galactic latitude, the high DM suggests an extragalactic origin.

None of the FRBs from Parkes or Arecibo to-date have been associated with a GRB, although the theoretical possibility that they might be connected has motivated several experiments to look for coincident GRB/FRB emission. \citet[][hereafter B12]{Bannister2012} performed a targeted survey to detect short duration radio pulses from GRBs at 1.4\,GHz. B12 observed nine GRBs, both long and short, using a 12\,m radio telescope located at the Parkes Observatory. B12 reported the detection of two single radio pulses from two long GRBs with S/N $>$ 6$\sigma$.  The pulses were detected 524 and 1076 seconds after the $\gamma$-ray emission from the GRBs. The DMs of the pulses were 195 pc $\rm cm^{-3}$ and 570 pc $\rm cm^{-3}$, which are large compared to the expected Galactic electron-density contribution \citep{Cordes2002} to the DM, indicating the possibility that the bursts are extragalactic in origin and possibly from the GRBs.

Similar to the B12 experiment, there have been a number of past and ongoing experiments searching for FRB-like emission from the prompt phase of GRBs at low and high radio frequencies \citep{Obenberger2014, Staley2013, Balsano1999,  Benz1998, Baird1975}. \citet{Baird1975} observed 19 GRB events at 151\,MHz between 1970 and 1973, with a time resolution of 0.3s; they searched for coincident  radio pulses with two or more stations over periods $\rm -1 hr$ to $\rm +10 hr$ relative to the GRB event times. Their search failed  to locate any astronomical radio pulse above their sensitivity limit of $\rm 10^{5} \, Jy$. Two other GRBs were observed by \citet{Dessenne1996}, who performed an automatic rapid radio follow-up experiment at 151\,MHz, with a time resolution of 1.5\,s; they reported a non-detection at a sensitivity of $\rm 3 \sigma$ of 73\,Jy. \citet{Benz1998} observed 7 GRBs between 1992 and 1994, covering a large range of frequencies between 40 and 1000\,MHz using three similar spectrometers, with a time resolution of 0.15\,s. Although the experiment sensitivity was low ($\rm 10^{5} \, Jy$), the spectral coverage made this experiment very interesting. Similarly, \citet{Balsano1999} observed 32 GRBs at 73.75\,MHz from 1997 to 1998 and had a wide range of sensitivity for each GRB. They reported a non detection at sensitivity limit of $\rm \sim 200 \, Jy$ for integration times of $\rm 50 \, ms$.

Most Recently, \citet{Obenberger2014} reported the observations of 12 GRBs at 74\,MHz, five GRBs at 52\,MHz, and 17 GRBs at 37.9\,MHz with a time resolution of 5\,s. Similarly, \citet{Staley2013} reported the observation of four GRBs at 15\,GHz with a time resolution of 0.5\,s. Both \citet{Staley2013} and \citet{Obenberger2014} did not detect any FRB-like emission from the observed GRBs.

The sensitivities of the above mentioned experiments were generally low, however they have motivated a number of follow-up experiments including this work and  a recent follow up project proposed by Bannister et al (2013) (VLBA project codes BB317 and BB325)\footnote{https://safe.nrao.edu/wiki/bin/view/VLBA/ProposalsAndDispositions} to use one or more Very Long Baseline Array (VLBA; \citet{Napier1994}) dishes to automatically follow up GRB alerts during telescope idle time. The VLBA-GRB experiment searches for FRBs using the existing V-FASTR system \citep{Randall2011}.

In this paper, we describe observations performed at 2.3\,GHz using the Hobart 26\,m radio telescope to search for FRBs associated with GRBs. We designed our experiment to be similar to the B12 experiment with some improvements. Our aim was to observationally test the possible association of FRBs with GRBs and compare our results with the B12 results. Section \ref{sec:observations} of this paper describes the observations performed. A detailed description of our single pulse search pipeline and verification is given in section \ref{sec:dataprocessing}. In section \ref{sec:ResultsandDiscussion} we analyse our results and compare them with the B12 results. Conclusions are presented in section \ref{sec:conclusion}.  
   
\section{Observations} 
\label{sec:observations}
\begin{sidewaystable}[htpf]
\begin{center}  

\caption{Details of five GRBs observed using the 26\,m radio telescope. $\rm P_{BAT}$ is the BAT position uncertainity, $\rm T_{BAT}$ is the time when BAT detected a GRB, $\rm T_{alert}$ is the time when the GCN alert notifications were sent, $\rm T_{90}$ is the GRB duration based on the BAT light curve, $\rm T_{rec}$ is the time when the recorders at 26 m Hobart telescope started recording the data from the source, $\rm T_{obs}$ is the observation duration with the radio telescope and $\rm T_{XRT}$ is the time when X-ray data were recorded by the XRT on Swift. \label{tbl-1}} 
\bigskip
\footnotesize\setlength{\tabcolsep}{2.5pt}	
\begin{tabular}{lllcllllllc}  

\tableline\tableline
Source & $\alpha$ & $\delta$ & $\rm P_{BAT}$ & Trigger  & $\rm T_{BAT}$  & $\rm T_{alert}$  & $\rm T_{90}$  & $\rm T_{rec}$ & $\rm  T_{obs}$ & $\rm T_{XRT}$\\
& (J2000.0) &  (J2000.0) & ($\rm arcmin$) & Date  &  (UT) & (UT) & (s) & (UT) &(s) & (UT + s)\\
\tableline 

GRB 111212A & $+20^{h}$ $41^{m}$ $37^{s}$  & $-68^{\circ}$ $37^{\prime}$ $01^{\prime\prime}$  & 3    & 2011-Dec-12 & 09:23:07 & 09:24:06 & 30.86  &  09:24:50  & 1200 & $\rm T_{BAT}$ + 2856 \\  
GRB 120211A &  $+05^{h}$ $51^{m}$ $00^{s}$  & $-24^{\circ}$ $45^{\prime}$ $32^{\prime\prime}$  & 3   & 2012-Feb-11 & 11:58:28 & 11:59:29 & 101	 & 12:03:10  & 1800 & $\rm T_{BAT}$ + 119.5\\ 
GRB 120212A &   $+02^{h}$ $52^{m}$ $25^{s}$ & $-18^{\circ}$ $01^{\prime}$ $29^{\prime\prime}$  & 3   & 2012-Feb-11 &	09:11:22.74  & 09:11:43 & 20.42 &  09:13:20 & 1800  & $\rm T_{BAT}$ + 2676 \\   
GRB $\rm 120218A^{\dagger}$ & $+21^{h}$ $19^{m}$ $02^{s}$  & $-25^{\circ}$ $27^{\prime}$ $00^{\prime\prime}$ & 3 & 2012-Feb-18 &  00:49:22.14	& 00:50:17 & 7 	& 00:52:10  & 1800 & -\\  
GRB 120224A & $+02^{h}$ $43^{m}$ $54^{s}$  & $-17^{\circ}$ $47^{\prime}$ $32^{\prime\prime}$  	& 2.4	& 2012-Feb-24 & 04:39:56.49 &04:40:40 & 2.06 & 04:42:00  & 1800 & $\rm T_{BAT}$ + 109.1 \\ 
\tableline 
\end{tabular}
\footnotetext{$\dagger$ No XRT light curve due to Sun observing constraints. Swift could not slew to the BAT position until 19:14 UT on 20-02-12. }
\end{center}
\end{sidewaystable}

We observed the five GRBs listed in Table \ref{tbl-1} at a central frequency of 2.276\,GHz, using the 26\,m radio telescope located at the Mount Pleasant Radio Observatory operated by the University of Tasmania\footnote{\scriptsize www.ra-wiki.phys.utas.edu.au} near Hobart, Tasmania.  A computer at the observatory is configured to receive GRB notifications via email from the Gamma-ray Coordination Network (GCN)\footnote{\scriptsize www.gcn.gsfc.nasa.gov}. The GCN system distributes the location of the GRBs detected by space-borne observatories such as Integral \citep{W2003}, Swift \citep{G2004}, Agile \citep{T2009} and Fermi \citep{A2009} in real time. A filter is applied to accept notices only from the Swift satellite, hence the GRBs listed in the Table \ref{tbl-1} are all observed by the Swift satellite. The gamma-ray Burst Alert Telescope (BAT) on Swift first detects the GRB and localizes the burst direction to an accuracy of $\rm \sim 3^{\prime}$. These are the co-ordinates used by the telescope control system. Given that the radio telescope's primary beam Full Width at Half Maximum (FWHM) for this experiment is about $\rm 0.35^{\circ}$, all the observed GRBs in this experiment fall within the beam. Only GRBs that are above the elevation limit ($12^{\circ}$) and within the visible range of the telescope ($-90^{\circ}$ $<$ $\delta$ $<$ $+30^{\circ} $) are observed. When an accessible GRB is identified, the telescope slews to the listed coordinates and then starts recording data. On average it takes approximately 140 seconds for the telescope to slew to the required location. 

Data were recorded using the Australian Long Baseline Array (LBA) Data Acquisition System (DAS)\footnote{\scriptsize www.atnf.csiro.au/technology/electronics/docs/lba\_das/lba\_das.html} \citep{phillips2009}. The LBA DAS has a sampler and  digital filter which receives two independent analog intermediate frequencies (IFs), the right and left circular polarised signals over a 64\,MHz bandwidth. The analog IFs are sampled and digitised to 8 bit precision and then re-sampled to 2 bits using digital filters \citep{phillips2009} before the digital representation of the voltage is recorded to computer disk. For each observation the system equivalent flux density (SEFD) of the telescope was measured to be 800-900\,Jy, which is calculated as $\rm SEFD=\frac{8 \: k_{B} \: T_{sys}}{\eta_{a} \: \pi \, D^{2}}$ where $\rm T_{sys}$ is the system temperature (92 $-$ 103\,K), $\rm k_{B}$ is the Boltzmann constant, $\rm \eta_{a}$ and $\rm D$ are the antenna efficiency (0.6) and the diameter of a circular aperture antenna (26\,m), respectively. The automatic gain control loops were turned off for the duration of the observations. The one sigma flux density sensitivity for the 26\,m telescope (to detect a pulse of width 25\,ms) is approximately 0.5\,Jy, calculated as $\sigma = \frac{\rm SEFD}{\sqrt{\rm \bigtriangleup{f} \: \bigtriangleup{t} \: n_{p}}}$ where $\bigtriangleup{f}$ and $\bigtriangleup{t}$ are the bandwidth (64\,MHz) and integration time (25\,ms), respectively, and $\rm n_{p}$ is the number of polarisations (2). The pointing error of the 26\,m telescope across the whole sky is $\backsim 21^{\prime\prime}$. 

The observations occurred over a period of 30 minutes for each GRB except for GRB 111212A which was observed for 20 minutes. Following observation of the GRB field, the telescope slewed to a nearby calibrator\footnote{\scriptsize LBA calibrators list- www.atnf.csiro.au/vlbi/observing/fringefinders.html} and  data were recorded for 20 minutes. A control observation was then undertaken by pointing the telescope at blank sky ($\sim 2^{\circ}$ away from the GRB position) for approximately 20 minutes and recording the voltage data.   
  
\begin{figure}[htp]  
\includegraphics[width=6in,height=6.0in]{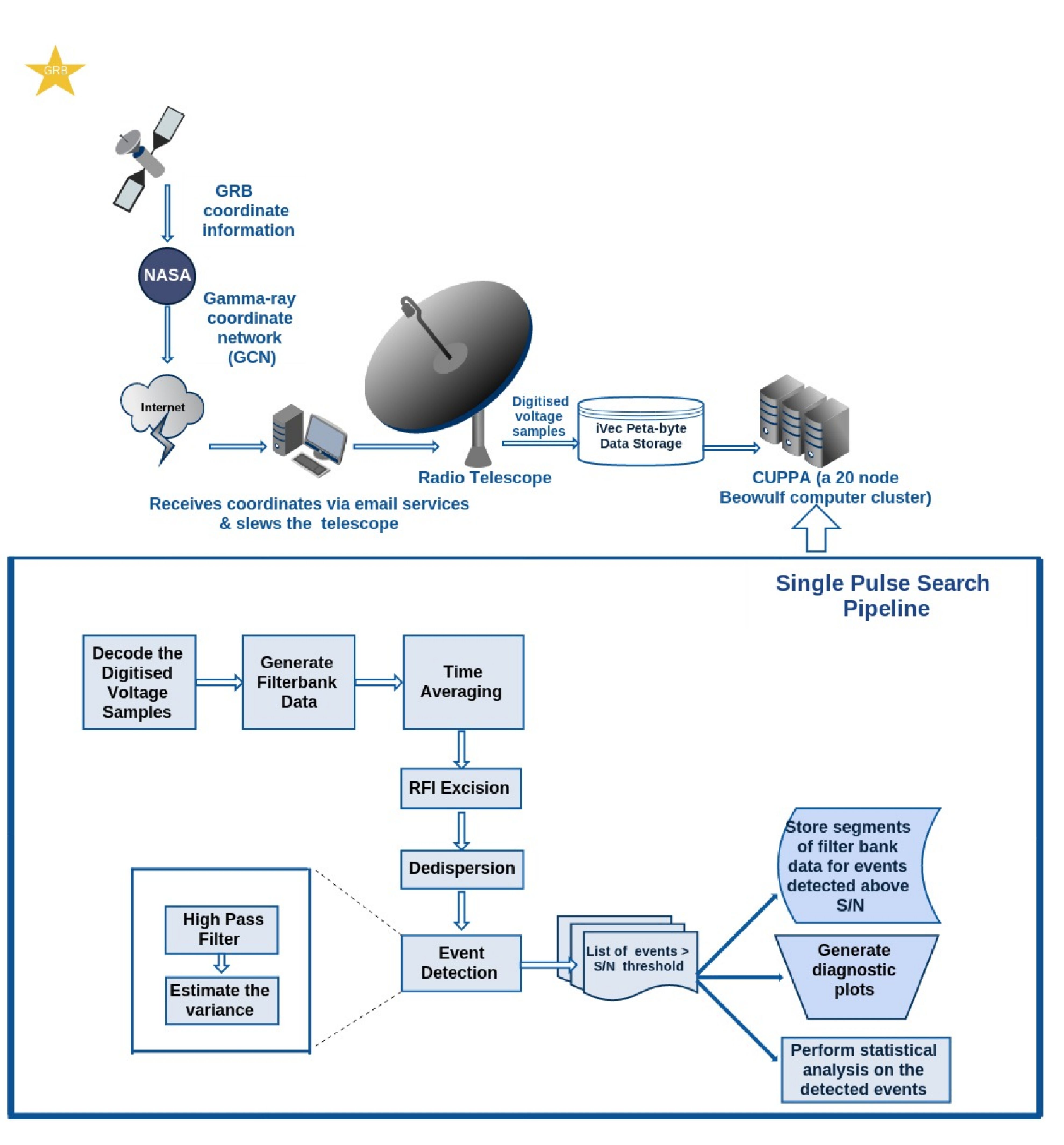}  
\caption{\scriptsize \textit{Top panel}: A schematic representation of the system where the space-borne GRB telescope sends the position information for GRBs to our ground-based radio telescope through GCNs. The radio telescope is then slewed to the GRB positions and data are recorded. The recorded data are then transferred to the PetaByte data storage facility at iVEC, in Perth and processed on a 20 node Beowulf computer cluster, CUPPA. \textit{Bottom panel}: Represents the single pulse search pipeline implemented in this paper, to search for FRBs from GRBs. The search pipeline has multiple stages: correlation; time averaging; radio frequency interference (RFI) excision; dedispersion; event detection and classification. In this experiment we generate time averages ranging from 640\,$\mu$s to 25.60\,ms (section 3.1.2). Each averaged time series go through the RFI excision, dedispersion, event detection and clasification stages independently. \label{flowchart}}  
\end{figure} 
  
\section{Processing and Verification}
\label{sec:dataprocessing}
The recorded data were transferred to the petabyte data storage facility at iVEC\footnote{\footnotesize www.ivec.org/about/}, in Perth. Data processing was performed on the Curtin University Parallel Processor for Astronomy (CUPPA), a 20 node computer cluster operated by Curtin University.

\subsection{Single Pulse Search Pipeline}
In this section we describe the data processing pipeline (Figure \ref{flowchart}) used to search for single short duration radio pulses in our data. The search pipeline has multiple stages: correlation; time averaging; radio frequency interference (RFI) excision; dedispersion; event detection; and classification of events. Several of our tools were reused from the V-FASTR experiment \citep{Randall2011} which can read LBA data natively.

\subsubsection{Correlation}
The first step in the processing pipeline was to pass the data through a spectrometer to form channelised total power for each polarisation. The 2-bit digitised voltage samples from each polarisation were decoded and sent to a software correlator. The correlator is an FX design with a configurable number of spectral channels and averaging time. We used $128 \, \times \,0.5$ MHz spectral channels, formed via FFT, and an averaging time of 640\,$\rm \mu$s. The output of the software correlator was written to files for subsequent processing.

\subsubsection{Time Averaging}

The intrinsic width of pulses are generally unknown. Because the dispersion and scattering affects the observed width of the pulse, a large parameter space of pulse widths must be searched \citep{cordes2003}. To obtain optimal sensitivity, the correlated time series were averaged for pulse widths ranging from 1.28\,ms to 20.48\,ms. This was achieved by averaging $2^{n}$ time samples together, where $\textit{n}$ ranges from 1 to 5, and searching each averaged time series independently. An additional time averaged series was generated at 25.60\,ms by averaging 40 time samples together (which is not a power of 2), to match the width used by B12 in their experiment.  

\subsubsection{RFI Excision}
Observations at 2.3\,GHz are generally affected by Radio Frequency Interference (RFI) from terrestrial transmitters, radars, satellites etc., which changes the total power as a function of frequency and time. Therefore, identification and excision of RFI was undertaken. 

\begin{figure}[htp] 
\center
\includegraphics[width=6.15in,height=2.5in]{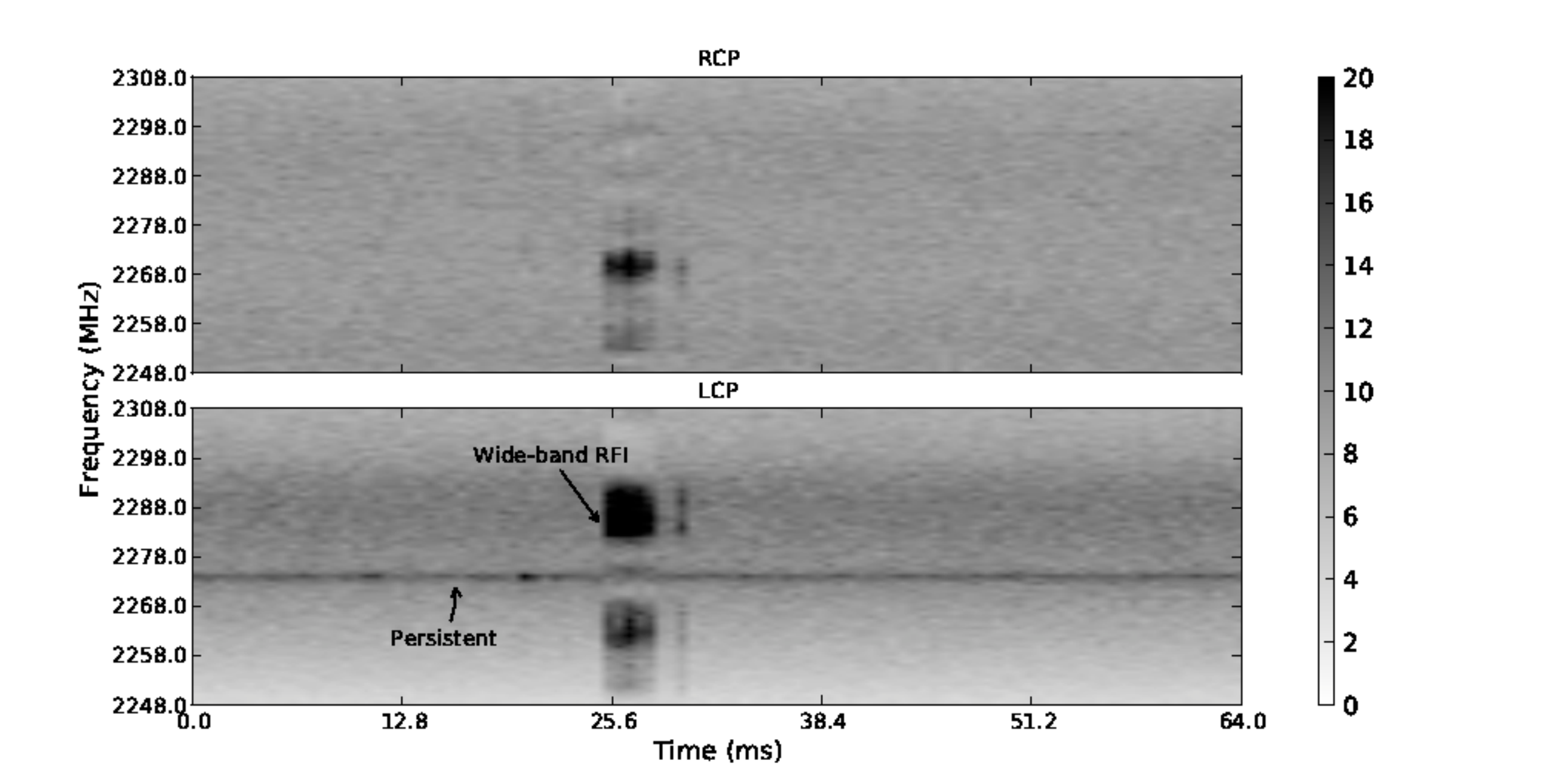}\\
\includegraphics[width=6.15in,height=2.5in]{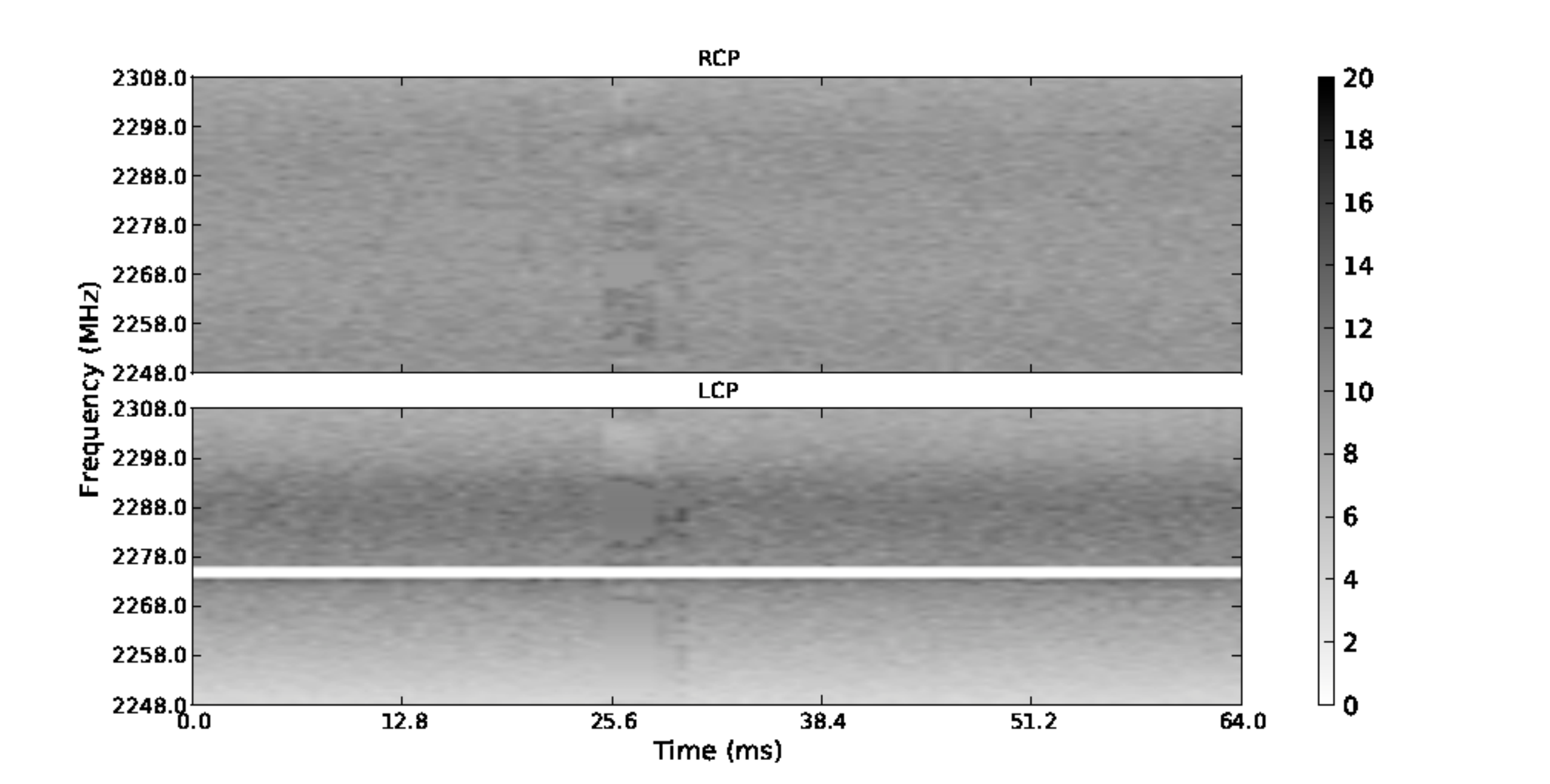}\\
\figcaption{\scriptsize Example dynamic spectrum shows RFI excision. Intensity is arbitrary power units. \textit{Top panel}: an example of a wide band RFI and persistent narrow band RFI. \textit{Bottom panel}: shows the results of RFI excision algorithm on the data  \label{RFI}}  
\end{figure}

\begin{figure}[htp] 
\center
\includegraphics[width=6.15in,height=2.1in]{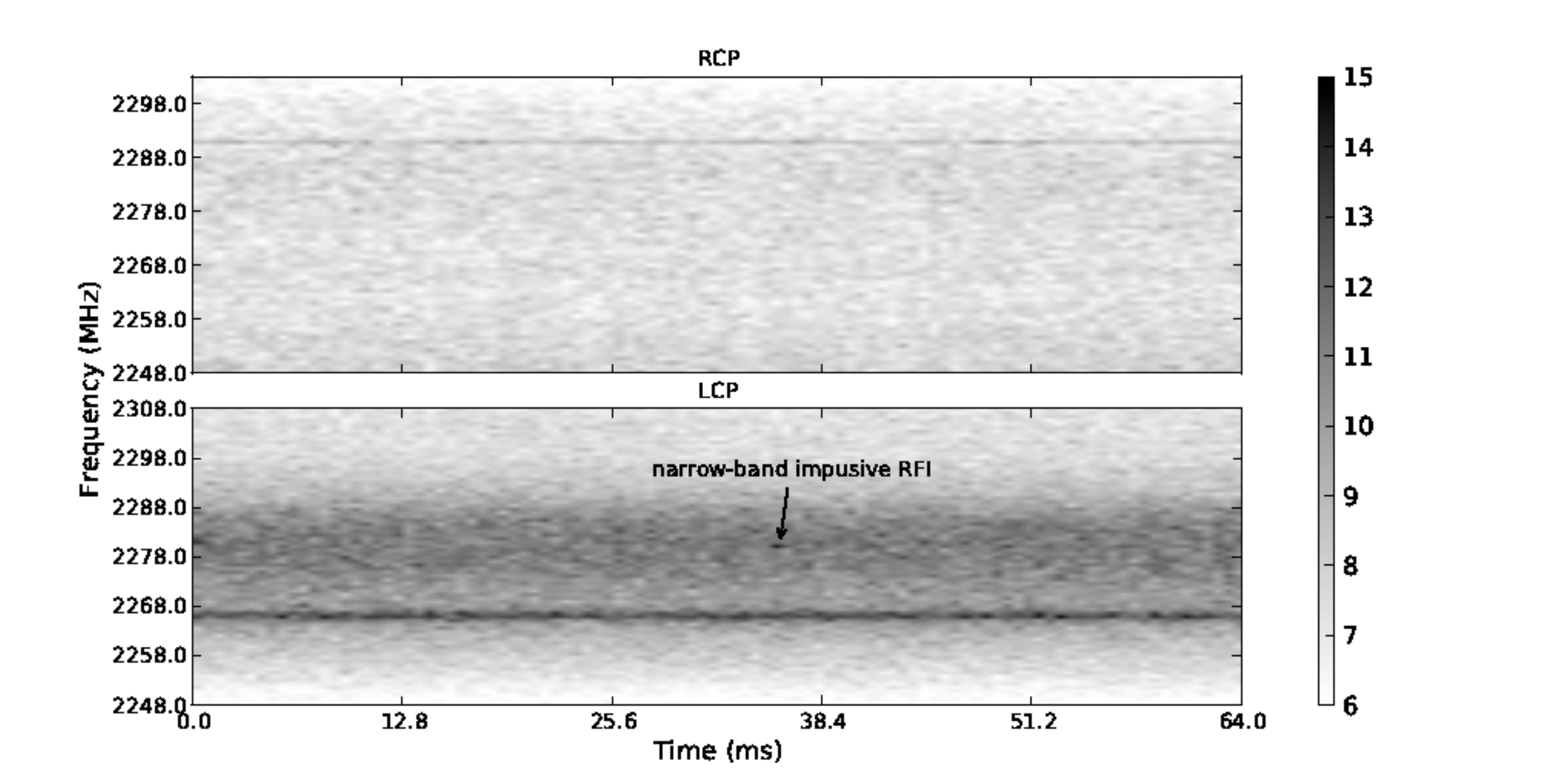}\\
\includegraphics[width=6.15in,height=2.1in]{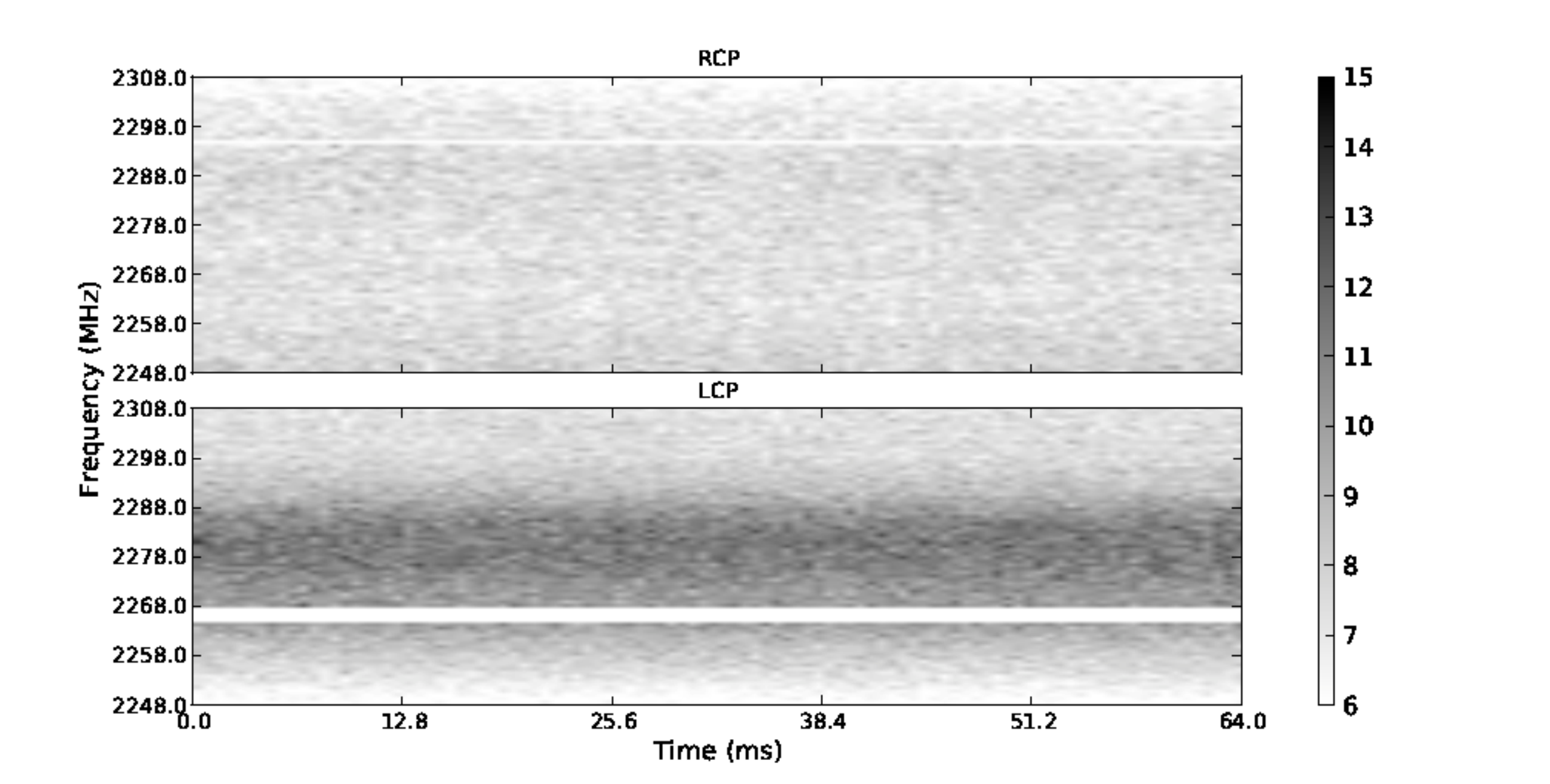}
\caption{\scriptsize Example dynamic spectrum shows RFI excision. Intensity is arbitrary power units. \textit{Top panel}: an example of a narrow band impulsive RFI. \textit{Bottom panel}: shows the results of RFI excision algorithm on the data.  \label{impulsiveRFI}}  
\end{figure}

First, we excise the time samples in each frequency channel that were affected by sporadic RFI. The top panel of Figure \ref{RFI} shows an example of sporadic, wide-band RFI, where the power level in all the frequency channels over short time scales (in this case $\sim$ 4\,ms) is much higher when compared to the typical mean power in neighbouring time samples across the frequency band. The top panel of Figure \ref{impulsiveRFI} shows an example of sporadic narrow-band RFI, where the power level in a frequency channel over a short time (in this case $\rm \sim 640 \, \mu$s) is high when compared to the typical mean power in the neighbouring time samples in that frequency channel. Our RFI excision algorithm used a thresholding method to identify samples in the time domain which were affected by RFI. First, we calculated the local variance ($\sigma_{l}^{2}$) of the time samples in each frequency channel over an adjustable time range ($\sim$ 10 sec) and then compared to the power (amplitude) of each time sample. The time samples that exceeded a pre-set threshold (here the threshold was set to $5 \sigma_{l}$) were excised from the time series and replaced with the mean power value estimated for each frequency channel over the band. We conservatively chose a 5$\sigma_{l}$ threshold for RFI excision, because we did not want to clip the thermal noise distribution\footnote{\scriptsize  Assuming the power level in the data follows the normal distribution, the probability of detecting an event above some significant threshold due to thermal noise fluctuations in the data is determined using the cumulative distribution function  (see section 4.1.1).  Thus the probability of detecting an event $> 5\sigma$ is $2.87 \times 10^{-7}$  and for $> 3\sigma$ the probability is $1.3 \times 10^{-3}$. We have $\sim 2 \times 10^{6}$ independent samples, therefore, at $5\sigma$ we would be effectively clipping $\sim$ 1 sample as a false positive, compared to when a 3$\sigma$ threshold is chosen we would be effectively clipping $\sim$2600  samples. This could skew the thermal noise distribution in the data.}. The bottom panels of Figures \ref{RFI} and \ref{impulsiveRFI} show the results of our excision algorithm.

Next, we manually identified and excised persistent narrow-band RFI which are constant in time over a frequency channel. The top panel of Figure \ref{RFI} shows an example of persistent narrow band RFI where the power in a frequency channel is higher compared to the power in the neighbouring channels across the frequency band. The affected channels, including the lower band edge spectral channel (DC channel generally has spurious high power due to sampler offsets), were recorded in a flagging file. The dedisperser (the next stage in the pipeline) reads the flagging files for bad channels and ignored the bad channels in further processing.  We repeated the RFI excision process on all the time averaged data series.

\subsubsection{Dedispersion}
The next step in the pipeline is dedispersion. Any incoming astronomical impulsive radio signal will, in general, be dispersed in time across the spectral channels due to the ionised component of the interstellar and intergalactic media \citep{Lbook2012}. The intrinsic signal can largely be recovered by incoherent dedispersion, by applying an appropriate time delay for each spectral channel and adding power across all the spectral channels. The time delay between two frequencies is given by 
$ \rm \Delta t = 4.15 \, ms \, \times \, DM  \, \times \, (\nu_{lo}^{-2} - \nu_{hi}^{-2})$, where the DM (pc $\rm cm^{-3}$) is the integrated column density of free electrons along the line of sight and $\nu_{lo}$ and $\nu_{hi}$ are the low and high frequencies in GHz, respectively \citep{Lbook2012}.

We use DART (Dedisperser of Autocorrelations for Radio Transients) to perform the incoherent dedispersion, a detailed description of which is given by \citet{Randall2011}. The incoming data to DART are ordered in frequency and time for each polarisation. The dedispersion process sums the power across channels, with the delay for each channel according to the frequency and DM. At the output, DART combines both polarisations and generates a dedispersed time series for each DM. We dedisperse the data with DM steps as listed in Table \ref{DMtable} ranging from 0-5000 pc $\rm cm^{-3}$.  The choice of DM steps was made by setting the time delay ($\rm \Delta t$) between highest ($\nu_{hi}$) and lowest ($\nu_{lo}$) spectral channels equal to the data sampling time (integration time). Choosing DM steps smaller than the data sampling time leads to highly correlated time samples, since the dedispersed time series are virtually identical for neighbouring DM steps. DART is flexible when compared to the Taylor tree algorithm \citep{Taylor1974}, because it can dedisperse the signals beyond the diagonal DM, the DM where the gradient of the dispersion curve is one time sample per spectral channel. The DM steps across the DM range are linearly distributed for the first few DM steps, before progressing geometrically, with a growth rate of 1.02. This results in equal spacings at lower DM and increasingly larger spacings at higher DM. Table \ref{DMtable} summarises the number of DM steps used for each time average.

\begin{table}[htp]   
\begin{center}  

\caption{\footnotesize For each time average, the table lists the number of DM steps used to dedisperse the data. \label{DMtable}} 
\footnotesize
\begin{tabular}{ll}  
\tableline\tableline
Time Averaging (ms) & $\rm N_{DM}$ \\
\tableline
0.64   & 149  \\
1.28   & 114  \\
2.56   &  79   \\
5.12   &  45   \\
10.24  &  23   \\
20.48  &  12   \\
25.60   &  9   \\
\tableline
\end{tabular}
\end{center}
\end{table}

\subsubsection{Event Detection}
The next step in the pipeline is detection. After dedispersion, each time series is searched for events above a defined signal to noise threshold. In the detector, the time series are divided into small time segments  ($\sim 1.3$ s). We chose this length of time because visual inspection of the data showed that the total power sometimes changed on time-scales longer than this. If longer time segments were used this would artificially raise the variance estimate. Each time segment is high-pass filtered, by subtracting a 100 timestep moving average from each sample. This reduces any long term power variations due to system temperature ($\rm T_{sys}$) changes - atmosphere, receiver response etc., which may bias variance estimates on time scales of 100s of samples, while retaining sensitivity to short time scale variations. We note that the high-pass filter window is longer than the expected pulse width of an astronomical event.

Each dedispersed time segment is then searched for single events $>$ 5$\sigma$. The noise level of the time series for each DM step is estimated using the standard deviation. We also tried the robust trimmed  estimator, interquartile range (IQR) \citep{Fridman2008}, to estimate the noise level for each DM step. The number of events detected $> 5\sigma$ from the IQR and standard deviation techniques did not differ. We adopt use the standard deviation because the behaviour of this statistic is well understood when a finite number of samples are used for its estimation. 

\subsubsection{Classification of Detected Events $> 5 \sigma$}
\begin{figure}[htp]
\center 
\includegraphics[width=3.75in,height=3in]{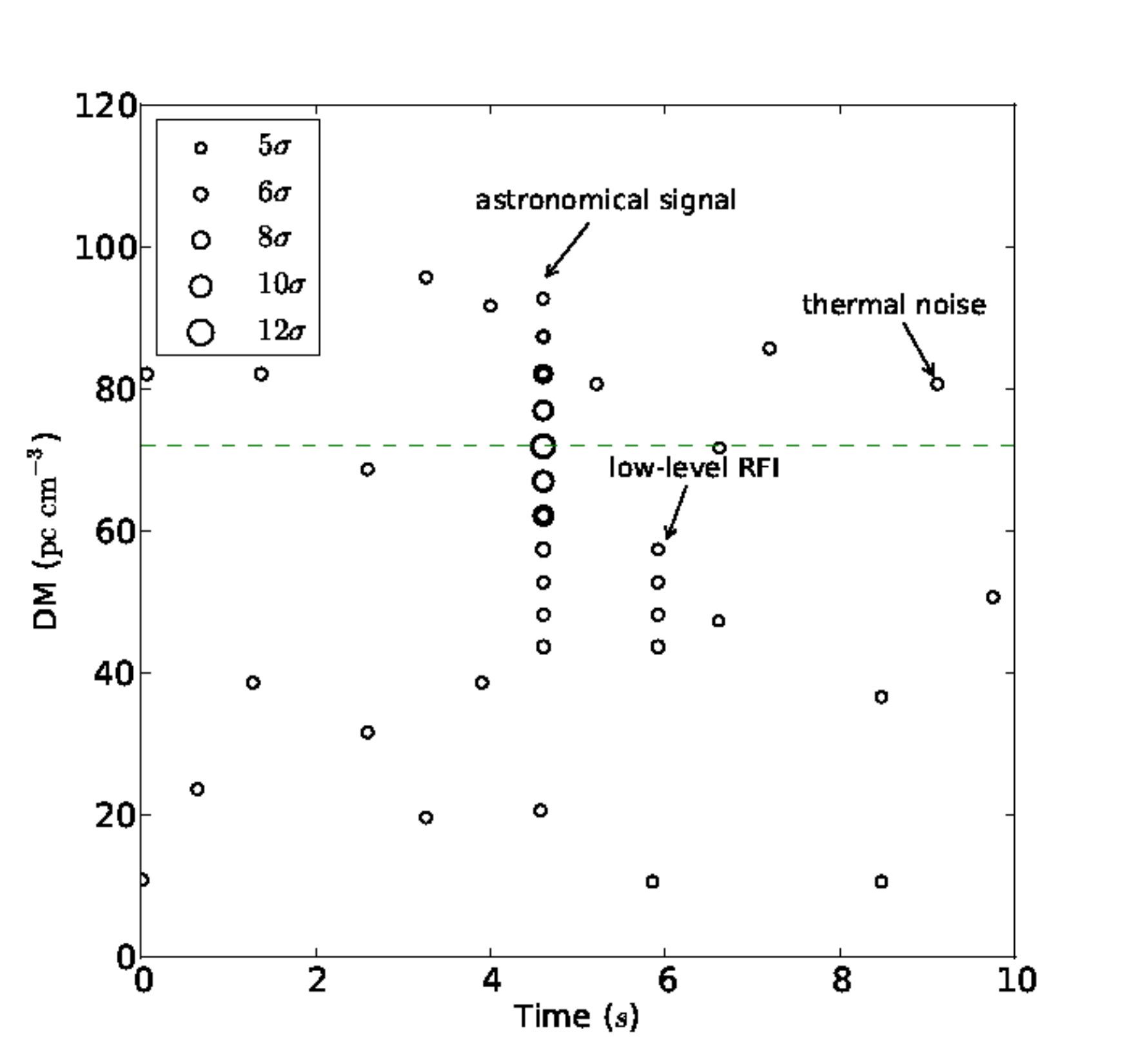}
\figcaption{\scriptsize An example diagnostic plot that demonstrates the difference between the signatures of astronomical signals,low level RFI and thermal noise fluctuations in the DM/time plane. The astronomical signal shows reduced S/N for DM steps away from the true DM (dashed line). The events due to thermal noise fluctuations appear at a single DM step. The events due to low level RFI are detected across multiple DM steps, but do not have the characteristic peaked shape that a bright pulse from an astronomical source would have. The size of the circles is proportional to the S/N. \label{DMtimeRFI} }
\end{figure}

The next step in the pipeline is the classification of detected  events  $> 5 \sigma$. Figure \ref{DMtimeRFI} demonstrates the differences between the signatures of astronomical signals, low level RFI and thermal noise fluctuations in the DM/time plane. An astronomical signal is expected to show reduced S/N for DM steps away from the true DM. The astronomical signals are broad band and any signals well above the minimum detection threshold are expected to be seen in multiple DM steps for a given time average and also in the consecutive time averages. Events due to thermal noise fluctuations will be seen at a single DM step and follow Gaussian statistics (most of the events would be close to the S/N threshold). Events due to low-level RFI will appear in multiple DM steps, but do not have the characteristic peaked shape that a bright pulse from an astronomical source would have.

The recorded events $> 5\sigma$ from our detection pipeline for all the time averages were visually examined for signatures of astronomical signals, low level RFI and thermal noise fluctuations using the diagnostic plots and were classified respectively. 

\subsection{Single Pulse Search Pipeline Verification} 
To test our single pulse search pipeline we used archival observations of the pulsar PSR J0835-4510 (the Vela pulsar). Observations were made using the same 26m radio telescope as used for our GRB observations.  The data recording system used was also identical to the recorder used for the GRB observations. Vela was observed at a central frequency of 1440 MHz with  right and left circular polarisations over a 64 MHz bandwidth. Vela has a steep spectral index, $\alpha = -1.2$ (S $\propto \nu^{\alpha}$), DM = 67.99 pc $\rm cm^{-3}$ and a period of 89.3 ms \citep{Manchester2005}. 

\begin{figure}[htp] 
\center
\includegraphics[width=4.75in,height=3.5in]{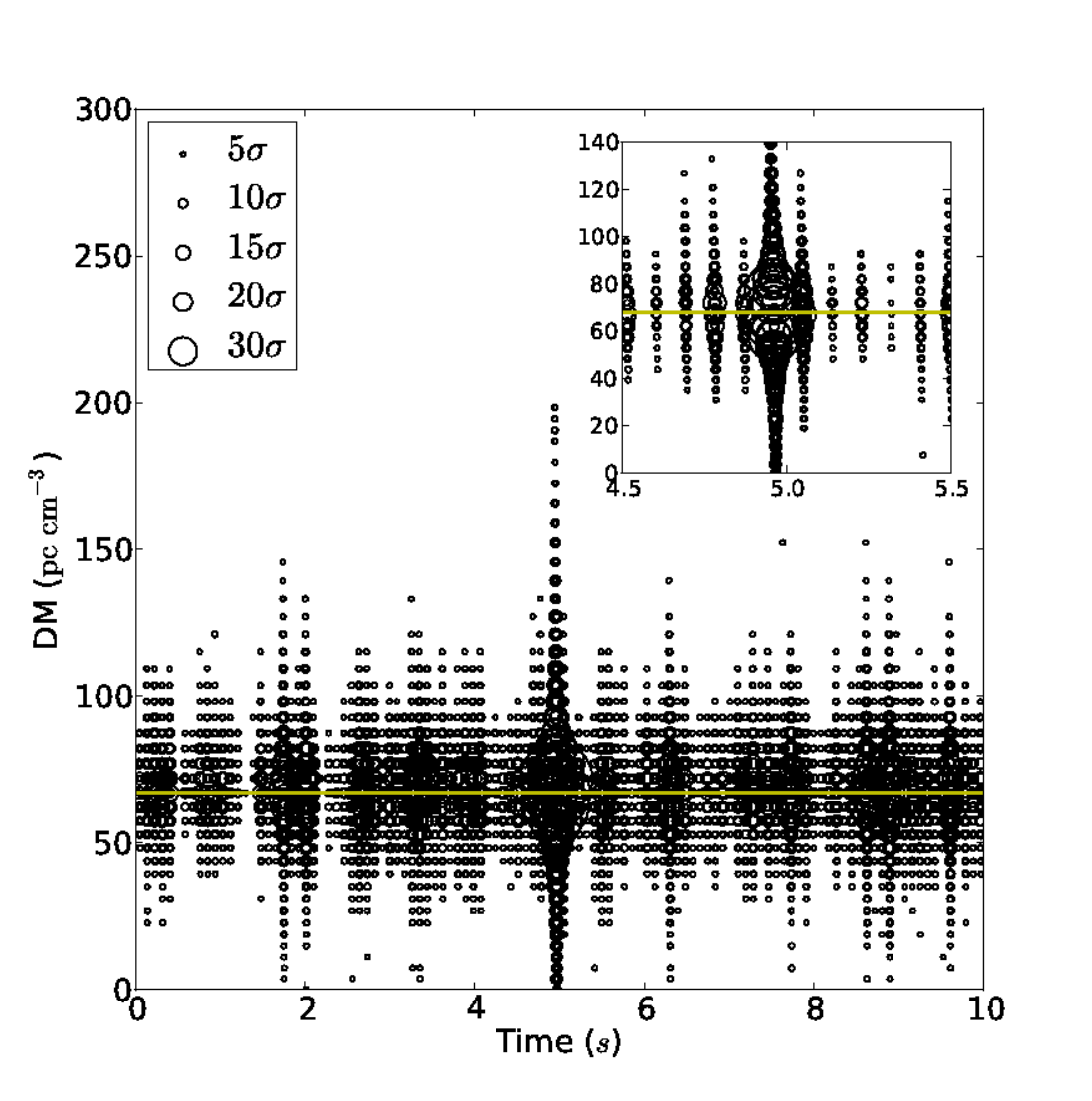}
\caption{\scriptsize Vela pulsar events detected $>$ 5$\sigma$ vs DM (pc $\rm cm^{-3}$) as a function of time. The size of the circles are proportional to the signal to noise ratio. Single pulses from Vela are detected at the true DM (horizontal line) and across multiple DM steps with reduced S/N. \label{DMTime}}  
\end{figure}

\begin{figure}[htp] 
\center 
\includegraphics[width=4.75in,height=3.35in]{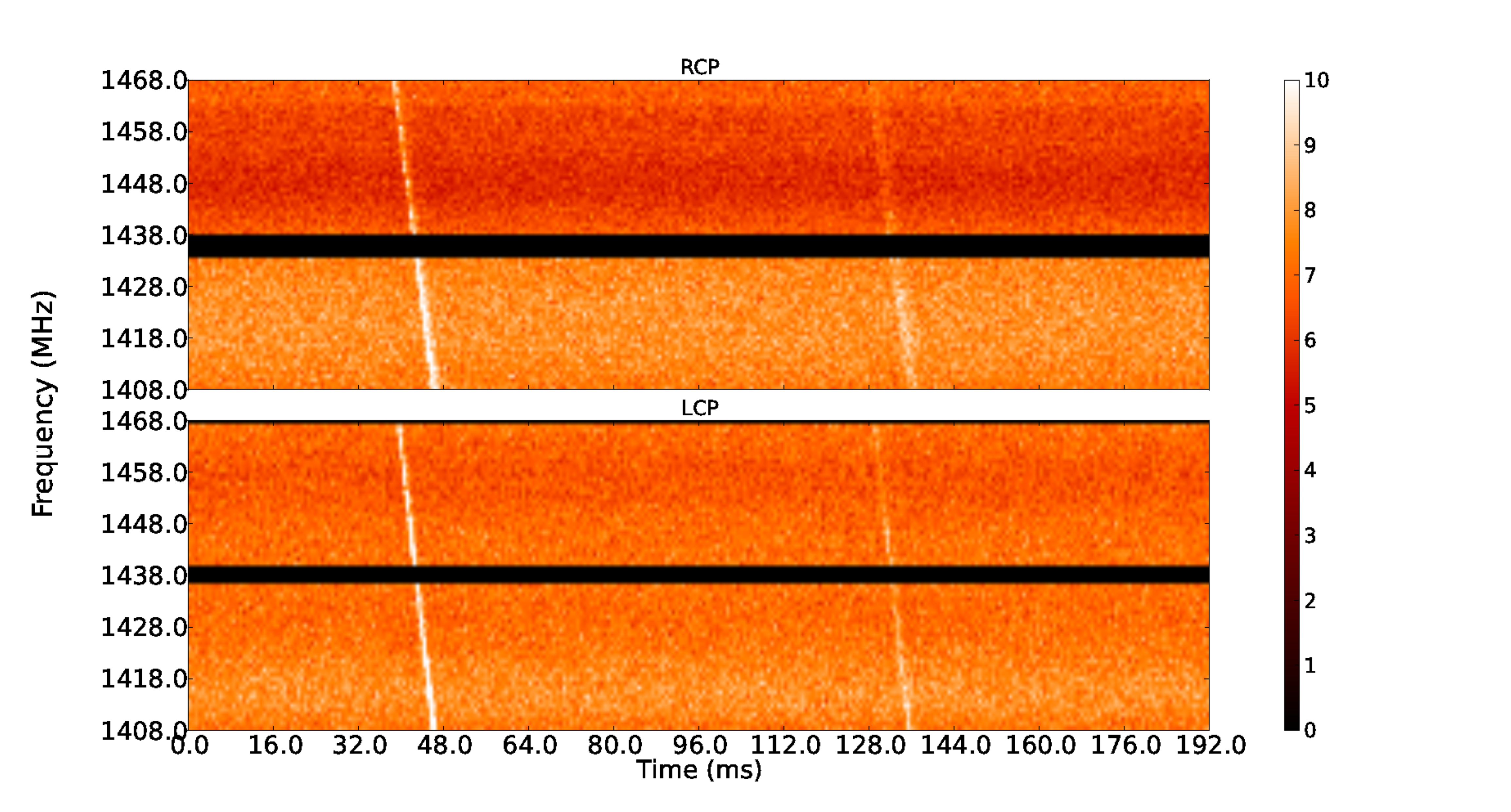}
\caption{\scriptsize Dynamic spectrum for Vela pulsar, time vs frequency. The colorbar represents the channel power in arbitary units. A strong pulse is seen at the left of the plot followed by a relatively weak pulse, correct period for separated by $\sim$ 89.3 ms which is the correct period for Vela. \label{Dynsplot}}  
\end{figure}

The recorded voltage data were correlated at 640 $\mu$s time resolution and 128 spectral channels, the same as for the GRB data. The correlated data were dedispersed over a DM range of 0-200 (pc $\rm cm^{-3}$), with 40 DM steps. The dedispersor output was searched for pulses above the S/N threshold of 5  at 640 $\mu$s time averaging and resulting events were examined to determine if they were real pulses from the pulsar in the time/DM plane (Figure \ref{DMTime}).  Individual pulses were detected at DM $\sim$ 67 pc $\rm cm^{-3}$  and at neighbouring DM steps with reduced S/N as expected. Segments of data were written out for each candidate pulse for further investigation. In the follow-up process, candidates were visually inspected via dynamic spectrum plots similar to those shown in Figure \ref{Dynsplot}. The dynamic spectrum shows individual dispersed pulses with a period of 89.3 ms  in time across the frequency band. Therefore, we have confirmed detection of individual pulses with Vela's period and DM via our single pulse search pipeline.     
  
\section{Results and Discussion} 
\label{sec:ResultsandDiscussion}
\subsection{Results}
\begin{figure}[htp] 
\center 
\includegraphics[width=5.75in,height=3.35in]{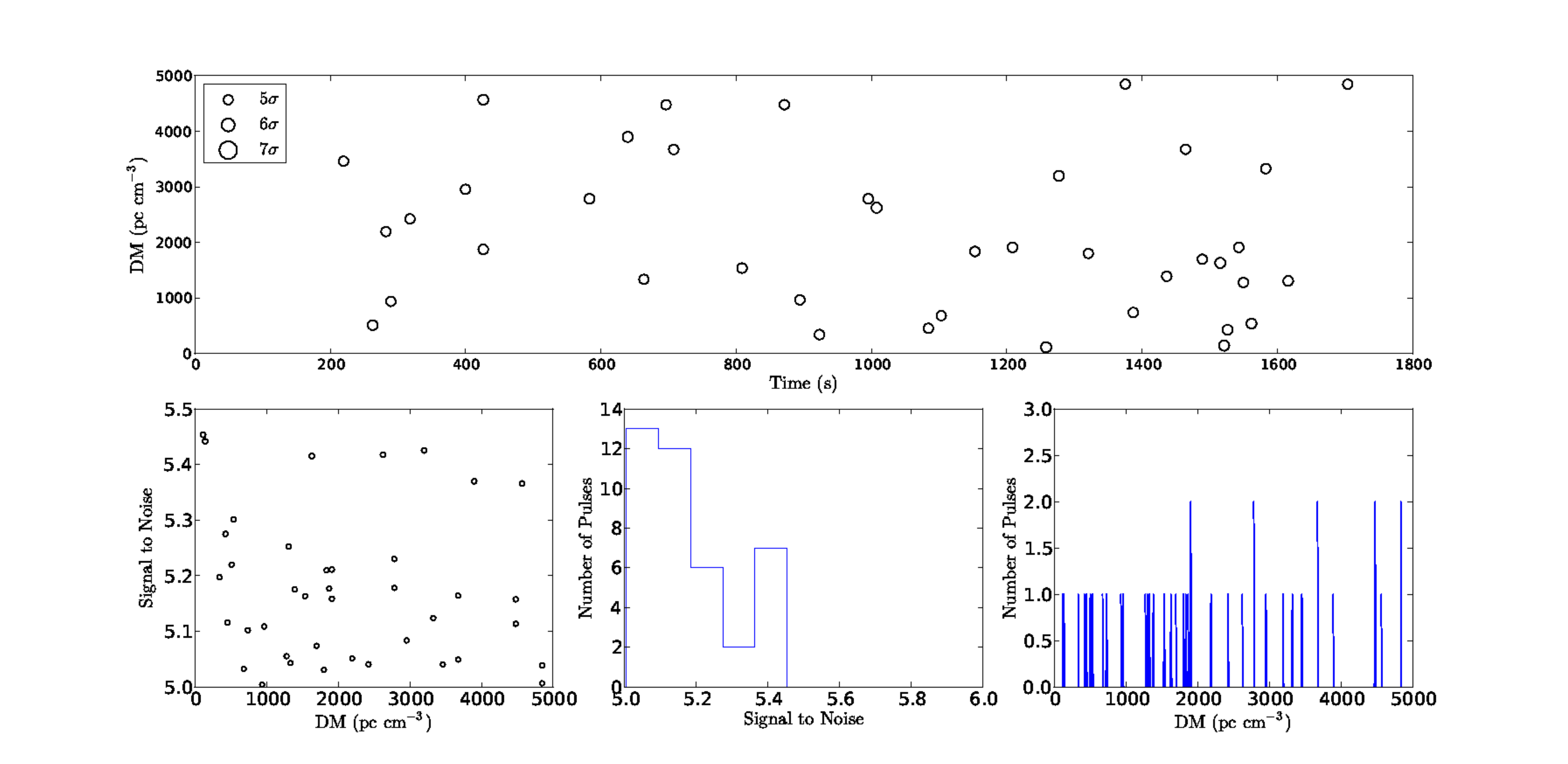}
\caption{\scriptsize Example diagnostic plot which illustrates the distribution of detected events $\rm > 5 \sigma$ for GRB 120211A at 1.28 ms time average. \textit{Top panel}: All events with S/N $\rm > 5 \sigma$ plotted vs DM and time. The size of circles are proportional to the S/N of the events. \textit{Lower left panel}: Scatter plot of the DM and S/N. \textit{Lower middle panel}: Distribution of the detected events $> 5 \sigma$. \textit{Lower right panel}: Number of detected events $> 5 \sigma$ vs DM.  \label{Diag}}  
\end{figure}

Here we discuss the results of our search for FRBs associated with GRBs using the single pulse search pipeline. The five GRBs have $\rm T_{90} > 2$ s based on the light curves from BAT. Processing the five data sets and visually inspecting the events via diagnostic plots did not show any signatures of astronomical signals. All the events detected by our detector were between $\rm 5 \sigma$  and  $\rm 6 \sigma$ and no events   $\rm \geq 6 \sigma$ were detected. Table\,\ref{tbl-2} lists the number of $\rm > 5\sigma$ events for all the GRBs and time averages, respectively. Figure \ref{Diag} shows an example diagnostic plot which illustrates the distribution of detected events $\rm > 5 \sigma$ for GRB 120211A at 1.28 ms time average.

In the absence of a characteristic signature of a astronomical signal $\rm > 6 \sigma$, we performed a statistical analysis on events between $\rm 5-6 \sigma$ to check if they were consistent with expectations  due to thermal noise fluctuations. This analysis gives us confidence that our search pipeline is working correctly and highlights the importance of sample variance in determining the significance of an event.

\subsection{Statistical analysis of the detected events $>$ $5 \sigma$}
\label{sec:stats}
Here we compare the number of events expected  due to thermal noise fluctuations to the number of events recorded by our pipeline. The number of events expected due to thermal noise fluctuations, $\rm N_{E}(I)$, above the detection threshold is given by,
\begin{equation}
\label{eq:one} 
\rm N_{E}(I > 5\sigma) = N_{t}\, N_{DM} \,  P(I > 5\sigma), 
\end{equation}  
where $\rm N_{DM}$ is the number of independent DMs, $\rm N_{t}$ is the number of independent time samples, and  $\rm P(I > 5\sigma)$  is the probability of an event above 5$\sigma$ due to thermal noise fluctuations (assuming the power fluctuation due to thermal noise follows the Gaussian distribution), and is given by the cumulative distribution function (CDF),
\begin{equation}
\rm  P(I > 5\sigma) = \int_{5\sigma}^{\infty} \mathcal{N}(\mu, \sigma^{2})  = \frac{1}{2} + \frac{1}{2} \rm erf\left[\frac{\mu - 5\sigma}{\sqrt{2}\sigma}\right],
\label{prob}
\end{equation} where erf is the error function and  $\mathcal{N}(\mu, \sigma^{2})$ is the Gaussian distribution with mean $\mu$ and variance $\sigma^{2}$.

\begin{figure}[htp]
\epsscale{0.6} 
\center 
\plotone{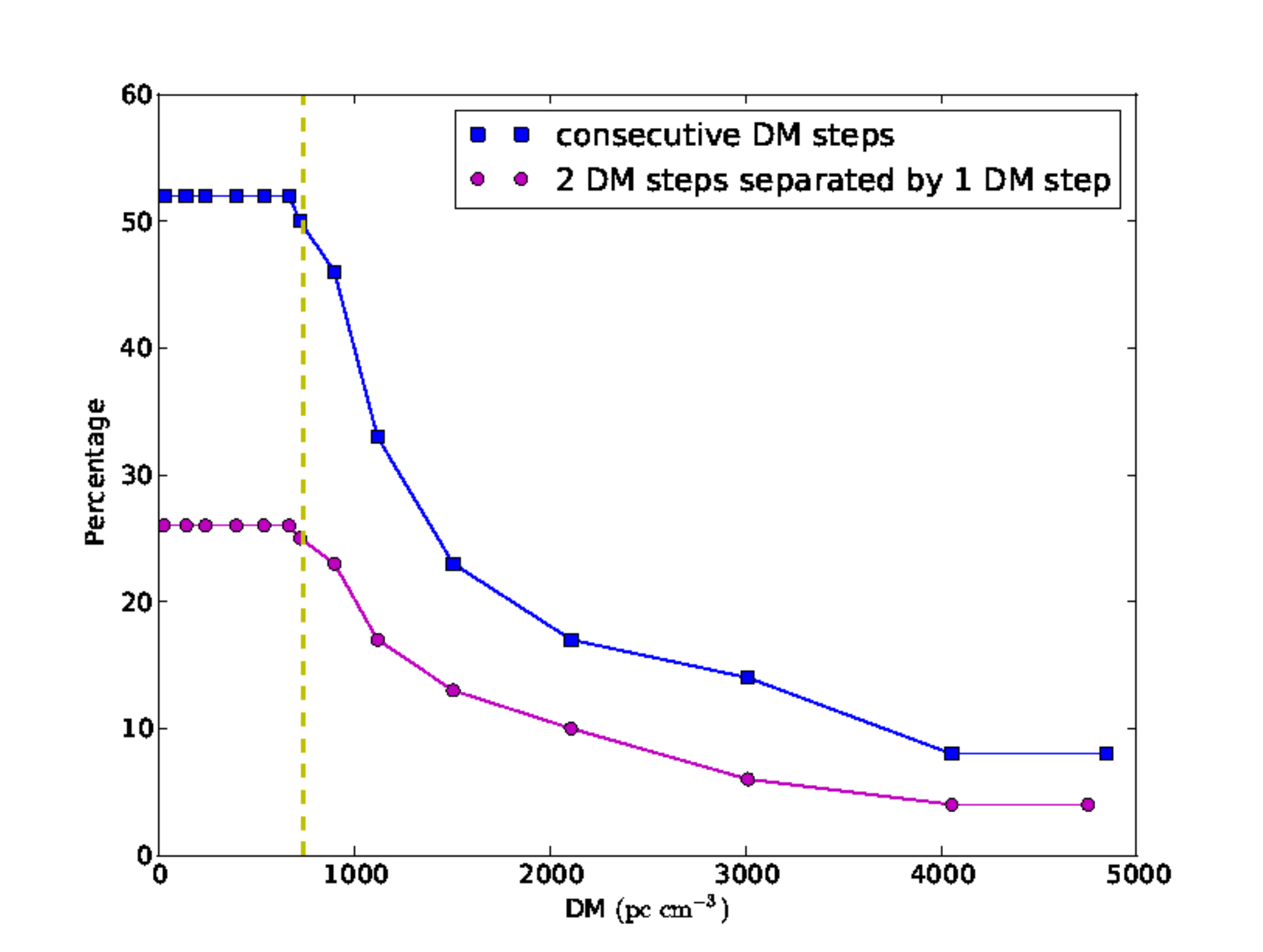}  
\figcaption{\scriptsize Percentage of identical power samples between consecutive DM steps at 640 $\rm \mu s$ time average. At lower DM (where DM steps are linearly distributed) $\sim$ 51- 52 $\%$ of power samples are identical between the consecutive DM steps and $\sim$ 26-27 $\%$ of power samples are identical between the two DM steps separated by one DM step. However, at the higher DM (where DM steps are exponentially distributed) the number of identical power samples between  DM steps reduces, which leads to less correlation in the dedispersed time series. The vertical dashed line represents the transition in the distribution of the DM steps from linear to exponential. \label{corr}}
\end{figure}

\subsubsubsection{Independent DM steps}
In Equation 1, we note that the $\rm N_{DM}$ steps should be independent. Given our choice of step progression we expect $\sim$ 50$\%$ of power samples to be identical between consecutive pairs at lower DM steps (Figure \ref{corr}). Here we perform an analytical calculation showing that the time series in each DM step at the lower end of the DM range can be treated as independent. 

Our calculation assumes that a 5$\sigma$ noise event is detected due to the addition of  power sample fluctuations in the time/frequency plane. Based on this assumption we consider the following scenarios: $\rm \left(i\right)$ The probability of two consecutive DM steps each having a $5\sigma$ event, since $\sim$ 50$\%$ of power samples are identical at the lower DM; and $\rm \left(ii\right)$ The probability of two DM steps separated by one DM step each having a $5\sigma$ event, since $\sim$ 25 $\%$ of power samples are identical at lower DM. In the former case, the probability is reduced by a factor of $\sim$ 100 compared with a single 5$\sigma$ event in a DM step due to thermal noise fluctuations. In the latter case, the probability indicates that events $> 5\sigma$ are consistent with events due to thermal noise fluctuations reduced by $\sim 10^{6}$ (a detailed description of the calculation is given in Appendix A). The small probability in the former and latter scenarios implies that  $5\sigma$ detections in two or more DM steps at the lower end of the DM range are independent.

\subsubsection{Uncertainties in the Estimation of $\rm N_{E}(I > 5\sigma)$} 
\label{subsec:Uncertainties}
The expected number of events due to thermal noise fluctuations, $\rm N_{E}(I > 5\sigma)$, is uncertain due to two effects: $\left(1\right)$ Events detected due to thermal fluctuations are stochastic and their number is characterised by the Poisson distribution; $\left(2\right)$ the variance measured from the data is a sample estimate, which has inherent uncertainty on the variance as it is being estimated over a finite number of samples. We take these effects into account in order to make our statistical analysis robust.

$\left(1\right)$ \textit{Stochasticity of 5$\sigma$ events}: We note that the thermal noise is stochastic in nature, inducing an uncertainty in the estimation of $\rm N_{E}(I > 5\sigma)$ which can be quantified using the Poisson distribution. The probability of observing X events, when $\lambda$ number of events is expected, is given by 
\begin{equation}
\rm Pois(X;\lambda) = \frac{\lambda^{X} \, \rm exp (-\lambda)}{X!}.
\end{equation}

For large $\lambda$, the Poisson distribution approximates the Gaussian distribution where
$\rm X \sim Pois(\lambda); \lambda \geqslant 15 \Rightarrow X \sim \mathcal{N}(\lambda, \lambda)$ \citep{Han1993}.
Therefore, for $\lambda \geqslant 15$ we use the Gaussian approximation and the uncertainty in the number of events fluctuates around the mean ($ \rm \lambda  = N_{E}(I > 5\sigma)$) with a standard deviation $\rm \sqrt{N_{E}(I > 5\sigma)}$. For $\rm N_{E}(I > 5\sigma) < 15$, the Gaussian approximation is invalid, and we use Monte Carlo simulations to estimate the uncertainties. This is described in section 4.2.2 

$\left(2\right)$ \textit{Sample variance in estimating true signal variance}: A finite number of samples (N = 2048 time samples at the intrinsic temporal resolution) is used to estimate the mean and the standard deviation of the data. The true mean and standard deviation of the data are not known. This estimate has an associated uncertainty known as sample variance, because only a finite number of samples is used. The standard error on the estimation of the mean ($\rm \bigtriangleup \hat{I_{0}}$) and variance ($\rm \bigtriangleup \hat{\sigma}^{2}$) for N independent samples from the true value is given by $\rm \bigtriangleup  \hat{I_{0}} = \frac{\sigma}{\sqrt{N}}$ and $\rm \bigtriangleup \hat{\sigma}^{2} = \sqrt{\frac{2}{N-1}} \, \sigma^{2}$ respectively. The errors $\rm \bigtriangleup \hat{I_{0}}$ and $\rm \bigtriangleup \hat{\sigma}^{2}$ are propagated into the expression for the cumulative probability (Equation 2) and the variance on the cumulative probability is obtained. For a detection threshold of $5\sigma$, the variance is given by $\rm \left( \Delta P\right)^{2}$,
\begin{equation}
\rm \left( \Delta P\right)^{2} = \frac{1}{2\pi} \, \rm exp(-5^{2}) \, \left(\frac{1}{N} + \frac{5^{2}}{N-1}\right),
\end{equation}
where N is the number of time samples. The sample size must be large enough to minimize the sample variance, but should not be so large that the mean changes with in the sample which would artificially increase the variance estimate.

The uncertainty in the expected number of events due to sample variance is propagated to an uncertainty in  $\rm N_{E}(I > 5\sigma)$ using equations 1 and 4,
\begin{equation}
\rm \Delta_{N} \equiv N_{DM} \, N_{t} \, \Delta P.
\end{equation}

The two uncertainties in the estimate of $\rm N_{E}(I > 5\sigma)$ are independent and Gaussian distributed (Section 4.2.1), and therefore can be combined in quadrature,
\begin{equation}
\label{eq:deltaE}
\rm \delta E^{2} = (\sqrt{N_{E}(I > 5\sigma)})^{2} \pm \Delta{^{2}_{N}}.
\end{equation}

\subsubsection{Estimating the significance of $\rm N_{A}(I > 5\sigma)$ in two regimes}
\label{subsec:MC}
In the previous section, we combined the two uncertainties associated with the estimation of $\rm N_{E}(I > 5\sigma)$ in quadrature, since both uncertainties obey Gaussian approximations. Here we would like to note again that the Gaussian approximation in case 1 is valid only for $\rm N_{E}(I > 5\sigma) \gtrsim 15$. We observe from the Table 3 and 4 that for higher time averages ranging  $\rm 5.12-25.5 \, m$s, $\rm N_{E}(I > 5\sigma) < 15$, the Gaussian approximation for the uncertainty discussed in the section \ref{subsec:Uncertainties} is no longer valid. Therefore, for higher time averages ($\rm 5.12-25.5 \, m$s), we can not combine both uncertainties in quadrature.

To estimate the significance $p$ of the observed events ($\rm N_{A}(I > 5\sigma)$), we split the calculations into two regimes: $\left( 1\right)$ At lower time averages ranging $640 \mu$s -  2.56\,ms were $\rm N_{E}(I > 5\sigma) \gtrsim 15$, we calculate $p$ via the Gaussian approximation with mean $\rm N_{E}(I > 5\sigma)$ and variance $\rm \delta E$, given by the equations \ref{eq:one} and \ref{eq:deltaE} respectively; $\left(2 \right)$ At higher time averages ($\rm 5.12-25.5 \, m$s) where $\rm N_{E}(I > 5\sigma) < 15$, we use results obtained via Monte Carlo simulation to estimate $p$. In both regimes, we interpret a $p > 0.05$ to be consistent with the null hypothesis which is that the observed events $\rm N_{A}(I > 5\sigma)$ are  consistent with noise.

In the low time average regime $p$ is given by,
\begin{equation}
p \rm = \frac{1}{2} - \frac{1}{2} \rm \, erf\left(\frac{N_{A}(I > 5\sigma) - N_{E}(I > 5\sigma)}{\sqrt{2} \, \delta E}\right).
\end{equation}

\subsubsubsection{Estimating Significance $p$ via Monte Carlo Simulations for Higher Time Averages}

To determine the significance of the observed events $\left( \rm N_{A}(I > 5\sigma)\right)$ at  higher time averages (when $\rm N_{E}(I < 5\sigma) \, < \, 15$), we use the Monte Carlo (MC) technique and perform $10^4$ simulations of random data sets. The simulations were performed on the time averages in the range $\rm 5.12 - 25.60 \, ms$. The results of the simulations automatically include the combined uncertainty due the thermal noise and sample variance.

Our program generates an array of Gaussian distributed random data sets of duration 30 minutes and $n$ DM steps (where $n$ is given in Table \ref{DMtable}). The array is then passed through the event detection pipeline (described in detail in Section \ref{sec:dataprocessing}). 

\begin{figure}[htp]
\center
\subfloat[5.12 ms]{\includegraphics[width=3.15in,height=2.85in]{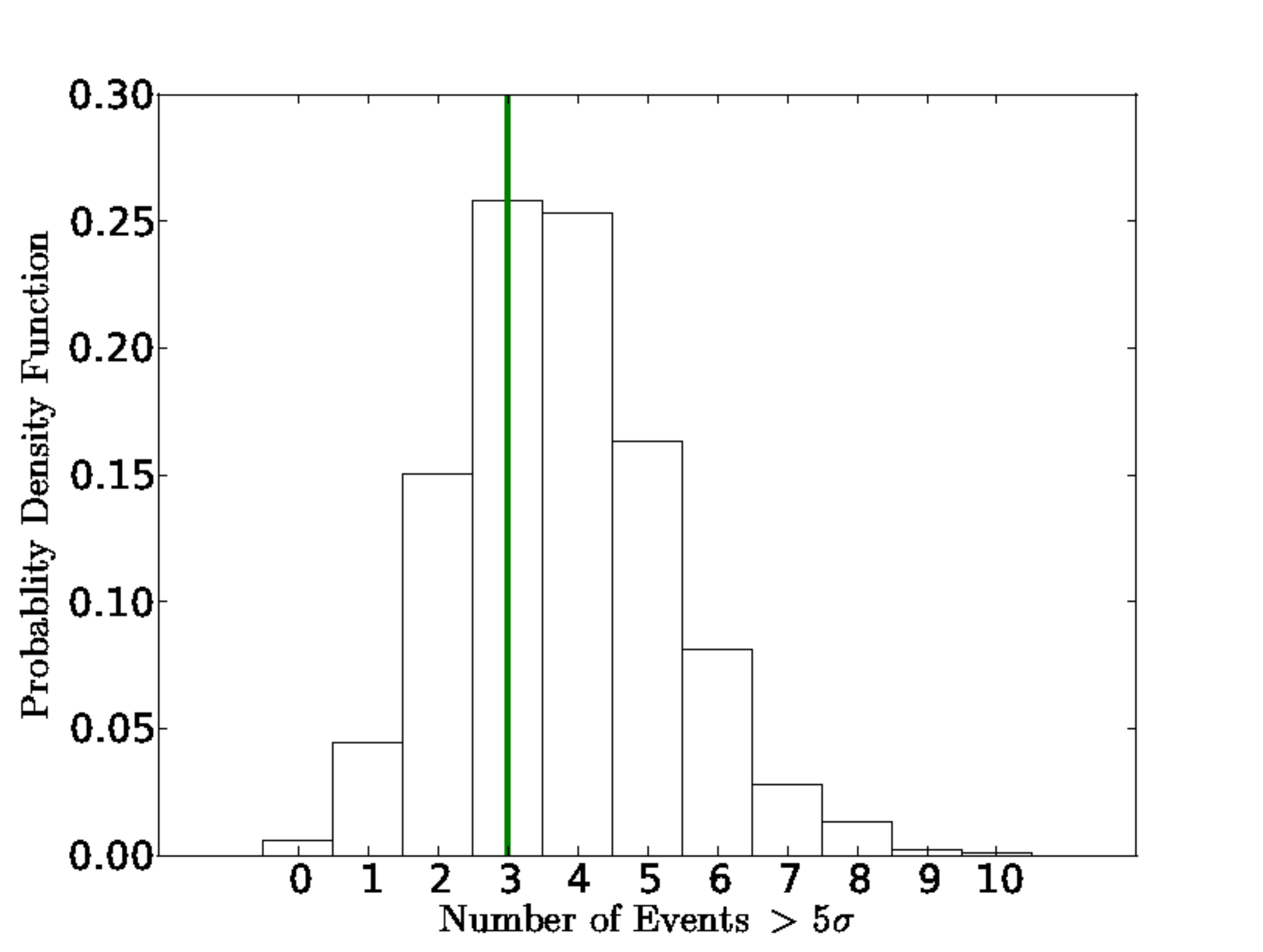}}
\subfloat[10.24 ms]{\includegraphics[width=3.15in,height=2.85in]{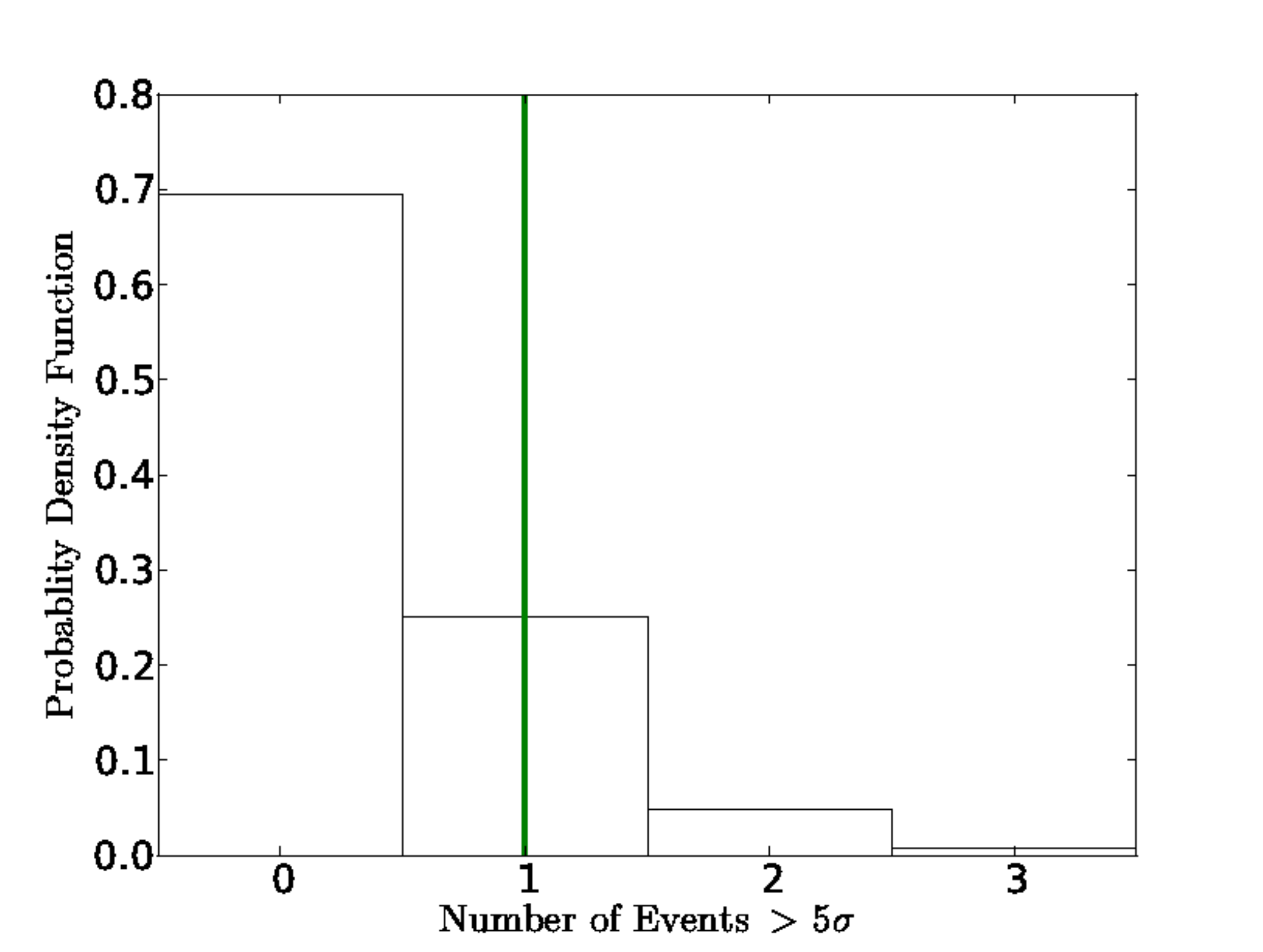}}
\caption{\scriptsize \textit{Left panel}: Example histogram obtained through MC simulations for GRB 120211A at 5.12 ms. The solid line in green represents the $\rm N_{A}(I > 5\sigma)$ for GRB 120211A at 5.12 ms time average. \textit{Right panel}: Example historgram obtained through MC simulations for GRB 120211A at 10.24 ms. The solid line in green represents the $\rm N_{A}(I > 5\sigma)$ for GRB 120211A at 10.24 ms time average \label{hist}.}
\end{figure}

For each iteration the MC simulation detects events $> 5 \sigma$. We performed $10^{4}$ iteration for each time average. Figure \ref{hist} shows an example histogram  for 5.12 and 10.28 ms time averages where we plot the events detected over $10^{4}$ iterations. The values in each bin are normalised to plot the probability density function, such that the integral over the range is unity. The significance ($p$), is determined by integrating the values in the histogram greater than and equal to $\rm N_{A}(I > 5\sigma)$ value for a given time average. For instance from Table \ref{tbl-2}, the number of events, $\rm N_{A}(I > 5\sigma)$, recorded for GRB120211A is three at 5.12 ms time resolution. The significance of $\rm N_{A}(I > 5\sigma)$ at the 5.12 ms time resolution can be calculated via the histogram summing the values greater than and equal to the bin number three (solid green line Figure \ref{hist}a). The region in the histogram greater than and equal to $\rm N_{A}(I > 5\sigma)$ value is termed the critical region\footnote{Critical region in statistics is the portion of a statistics bloc where a null hypothesis is rejected via the results of a test of the hypothesis.} in statistics. Integrating the point probabilities greater than  and equal to the critical value gives the significance $p$ \citep{kay1998}.

$\rm N_{E}(I > 5\sigma)$ for each higher time average is given by the mean of the distribution, which is calculated  by taking the average of values obtained through MC simulations (dotted lines in figure \ref{hist}). Here the duration of the random data sets was chosen to be the same as the observation duration for each GRB and blank sky data set.

\subsubsection{Applying the Statistical Analysis on the Obtained Results}

The statistical analysis was applied to all the GRBs for all averaging times. 
Table \ref{tbl-2} presents the results of the analysis. In all cases $p \rm > 0.05$, indicating that our recorded events are consistent with thermal noise fluctuations. In the absence of a characteristic signature of an astronomical pulse $\rm > 6 \sigma$, we conclude that the events  $> 5 \sigma$ are consistent with the expectations due to thermal noise fluctuations. There were no events with a statistically significant S/N.
 
Another approach to determine the probability of an event occurring due to thermal noise fluctuations is to measure the false positive rate when there is no GRB in the beam. Observing a patch of blank sky after the GRB observation preserves the noise properties and RFI information, assuming that no source of astronomical impulsive radio emission is in the field of view. We analysed the data from the blank sky observations, which were processed in a manner identical to that described for the GRB data.  Table \ref{tbl-3} shows the results of our analysis for the blank sky datasets. Again, in all the cases $p \rm > 0.05$ indicating that our recorded events are consistent with thermal noise fluctuations. 

\begin{table}[htp]
\tiny\addtolength{\tabcolsep}{-1pt}
\renewcommand{\arraystretch}{0.75} 
\caption{\footnotesize A summary of the statistical analysis performed on all the GRB datasets. The subtables (a), (b), (c), (d), (e) list the time averages used in this experiments. The columns are: the source name, $\rm N_{E}(I > 5\sigma) $ is the expected number of events above the S/N threshold due to thermal noise fluctuations, $\rm N_{A}(I > 5\sigma)$ is the actual number of events detected above the S/N threshold, $\rm  \sqrt{N_{E}(I > 5\sigma)}$ and $ \Delta{_{N}}$ are the uncertainites on the $\rm N_{E}(I > 5\sigma)$, $\rm \delta E$ represents the uncertainities combined in quadrature, $p$ is the significance where for $p > 0.05$ indicates that events $\rm > 5\sigma$ are consistent with events due to thermal noise fluctuations. The significance $p$ for the time averages at and above 5.12 ms were calculated via MonteCarlo simulations as described in section \ref{subsec:MC}. $\rm N_{E}(I > 5\sigma)$ for the time averages at and above 5.12 ms were estimated by taking the mean of the PDF in the histograms. \label{tbl-2}}  
\centering  
\resizebox{8.00cm}{!} {%
\footnotesize
 
\begin{tabular}{l c c c c c c}  
\hline
\\
GRB Name & $\rm N_{E}(I > 5\sigma) $ & $\rm N_{A}(I > 5\sigma)$ & $\rm \sqrt{N_{E}(I > 5\sigma)}$  & $\rm \Delta_{N} $ & $\rm \delta E$ &  $p$ \\
\\
\hline\hline 
\begin{tabular}{c}
\\
(a) 640 $\rm \mu$s \\
\\
\end{tabular}
\\
\hline
\\
GRB $\rm 111212A^{\dagger}$   & 83   & 85  & 9  & 49 & 49 & 0.48\\ 
GRB 120211A   & 125   & 120 	 & 11   & 70  & 70 & 0.53 \\
GRB 120212A   & 125   & 125   & 11  & 70  & 70  & 0.50 \\
GRB 120218A   & 125   & 129   & 11  & 70  & 70  & 0.48 \\
GRB 120224A   & 125   & 128  & 11   & 70  & 70  & 0.48 \\ [1ex]      

\hline
\begin{tabular}{c }
\\
(b) 1.28 $\rm ms$ \\
\\
\end{tabular}
\\
\hline\hline
\\
GRB $\rm 111212A^{\dagger}$ & 32   &  26	& 6 & 36 &  36& 0.56 \\ 
GRB 120211A   & 48   & 42           & 7 & 54      & 54 & 0.54 \\
GRB 120212A   & 48   & 48           & 7 & 54      & 54 & 0.50 \\
GRB 120218A   & 48   & 40           & 7 & 54      & 54 & 0.55 \\
GRB 120224A   & 48   & 51           & 7 & 54      & 54 & 0.48 \\ [1ex]      
\hline
\begin{tabular}{c }
\\
(c) 2.56 $\rm ms$ \\
\\
\end{tabular}
\\
\hline\hline
\\
GRB $\rm 111212A^{\dagger}$  & 11    & 12 & 3 & 18 & 18   & 0.48 \\ 
GRB 120211A   & 16   & 12 	 & 4 & 26  & 26    			 &0.56 \\
GRB 120212A   & 16   & 22     & 4 & 26 &   26 			  & 0.41 \\
GRB 120218A   & 16   & 18     & 4& 26  &  26 			  & 0.47 \\
GRB 120224A   & 16   & 22     & 4& 26   &  26           & 0.56 \\ [1ex]      
\hline
\begin{tabular}{c }
\\
(d) 5.12 $\rm ms$ \\
\\
\end{tabular}
\\
\hline\hline
\\
GRB $\rm 111212A^{\dagger}$  & 3    & 1  & -  &- & -   & 0.72 \\ 
GRB 120211A   & 4   & 3 	    & -  &- & -   &0.80 \\
GRB 120212A   & 4   & 3     & -  &- & -   & 0.80 \\
GRB 120218A   & 4   & 1     & - &- & -     & 0.99 \\
GRB 120224A   & 4   & 4     & - &- & -    & 0.54\\ [1ex]      
\hline
\begin{tabular}{c}
\\
(e) 10.24 $\rm ms$ \\
\\
\end{tabular}
\\
\hline\hline
\\
GRB $\rm 111212A^{\dagger}$  & 0    & 1	 & - &- & -     & 0.21 \\ 
GRB 120211A   & 0   & 1 	 & -  &- & -      &0.30\\
GRB 120212A   & 0   & 0     & - &- & -    & 1.00 \\
GRB 120218A   & 0   & 0     & - &- & -     & 1.00 \\
GRB 120224A   & 0   & 0     & - &- & -    & 1.00\\ [1ex]      
\hline 
\begin{tabular}{c}
\\
(e) 20.48 $\rm ms$ \\
\\
\end{tabular}
\\
\hline\hline
\\
GRB $\rm 111212A^{\dagger}$  & 0    & 0	 & - &- & -   & 1.00 \\ 
GRB 120211A   & 0   & 0 	    &  - &- & -    &1.00 \\
GRB 120212A   & 0   & 0     & -  &- & -  & 1.00 \\
GRB 120218A   & 0   & 0     & -  &- & -    & 1.00 \\
GRB 120224A   & 0   & 0     & -   &- & -  & 1.00\\ [1ex]      
\hline 
\begin{tabular}{c}
\\
(e) 25.6 $\rm ms$ \\
\\
\end{tabular}
\\
\hline\hline
\\
GRB $\rm 111212A^{\dagger}$  & 0    & 0	 & - &- & -   & 1.00 \\ 
GRB 120211A   & 0   & 0 		 & - &- & -    &1.00 \\
GRB 120212A   & 0   & 0     & -  &- & -    & 1.00\\
GRB 120218A   & 0   & 0     & -  &- & -   & 1.00 \\
GRB 120224A   & 0   & 0     & -  &- & -    & 1.00\\ [1ex]      
\hline 
\tablecomments{\footnotesize
 $\dagger$ GRB 111212A was observed over a period of 20 minutes.}
\end{tabular}}
\end{table}

\begin{table}[htp]
\tiny\addtolength{\tabcolsep}{-1pt}
\renewcommand{\arraystretch}{0.75} 

\caption{\footnotesize A summary of the statistical analysis performed on all the blank sky datasets. The subtables (a), (b), (c), (d), (e) list the time averages used in this experiment. The columns are same as for Table 2 \label{tbl-3} }   
\centering  
\resizebox{8.00cm}{!} {%

\footnotesize
\begin{tabular}{l c c c c c c}  
\hline
\\  
GRB Name & $\rm N_{E}(I > 5\sigma) $ & $\rm N_{A}(I > 5\sigma)$  & $\rm \sqrt{N_{E}(I > 5\sigma)}$  & $\rm \Delta_{N} $ & $\rm \delta E$ & $p$ \\
\\
\hline\hline 
\begin{tabular}{c}
\\
(a) 640 $\rm \mu$s \\
\\
\end{tabular}
\\
\hline
\\
GRB $\rm 111212A^{a}$   & 15      &15     & 4 & 12   &12 & 0.52 \\ 
GRB 120211A  		    & 83     & 83      & 8 &  49  & 49 & 0.50\\
GRB $\rm 120212A^{b}$   & 83      & -	  & -  &        -  & -        & -             \\
GRB 120218A  		    & 83     & 81     & 8 & 49  & 49 & 0.52\\
GRB 120224A             & 83      & 90      & 8 & 49 & 49  & 0.44\\ [1ex]      
\hline
\begin{tabular}{c }
\\
(b) 1.28 $\rm ms$ \\ 
\\
\end{tabular}
\\
\hline\hline
\\
GRB $\rm 111212A^{a}$    & 8      & 1       & 3 & 9    & 9 &0.77\\ 
GRB 120211A  		     & 32     & 	28	   & 6 & 36    & 36 &0.54\\
GRB $\rm 120212A^{b}$    & 32      & -      & -   &      -    &  -    & -            \\
GRB 120218A  		     & 32     & 35      & 6 & 36   & 36 &0.48\\
GRB 120224A              & 32     & 37       & 6 & 36   & 36 & 0.44\\ [1ex]      

\hline
\begin{tabular}{c }
\\
(c) 2.56 $\rm ms$ \\
\\
\end{tabular}
\\
\hline\hline
\\
GRB $\rm 111212A^{a}$    & 3      & 1        & 1 & 4      & 4&0.68\\ 
GRB 120211A  		     & 11     & 9	     & 3 & 18     &  18  &0.54\\
GRB $\rm 120212A^{b}$    & 11      & -        & -     & - &  -  & -            \\
GRB 120218A  		     & 11     & 7        & 3 & 18    & 18 &0.58\\
GRB 120224A              & 11     & 9        & 3 & 18    & 18 &0.54\\ [1ex]      

\hline
\begin{tabular}{c }
\\
(d) 5.12 $\rm ms$ \\
\\
\end{tabular}
\\
\hline\hline
\\
GRB $\rm 111212A^{a}$    & 0      & 0       & -    & - &-  & 1.00\\ 
GRB 120211A  		     & 3     & 5		   & -   &- & -  & 0.14\\
GRB $\rm 120212A^{b}$    & 3      & -       & -  &   -    &   -       & -            \\
GRB 120218A  		     & 3     & 2       & -  &- & -    & 0.55\\
GRB 120224A              & 3     & 2       & -   &- & -   & 0.55\\ [1ex]      
\hline
\begin{tabular}{c }
\\
(e) 10.24 $\rm ms$ \\
\\
\end{tabular}
\\
\hline\hline
\\
GRB $\rm 111212A^{a}$    & 0      & 0   & -    & - & - & 1.00\\ 
GRB 120211A  		     & 0     & 0	   &  -   & - & - & 1.00\\
GRB $\rm 120212A^{b}$    & 0      & -   & -  & -   & -  & -            \\
GRB 120218A  		     & 0     & 0    & -   &- & - & 1.00\\
GRB 120224A              & 0     & 0    & -   & - & -& 1.00\\ [1ex]      
\hline 
\hline
\begin{tabular}{c }
\\
(e) 20.48 $\rm ms$ \\
\\
\end{tabular}
\\
\hline\hline
\\
GRB $\rm 111212A^{a}$    & 0      & 0       & -   & - & -  & 1.00\\ 
GRB 120211A  		     & 0     & 0		   & -    & - & -& 1.00\\
GRB $\rm 120212A^{b}$    & 0      & -    & -  & - & -   & -            \\
GRB 120218A  		     & 0     & 0       & -    &- & - & 1.00\\
GRB 120224A              & 0     & 0       & -   &- & - & 1.00\\ [1ex]      
\hline
\begin{tabular}{c }
\\
(e) 25.6 $\rm ms$ \\
\\
\end{tabular}
\\
\hline\hline
\\
GRB $\rm 111212A^{a}$    & 0      & 0       & -    &- & - & 1.00\\ 
GRB 120211A  		     & 0     & 0	     	&  -  & - & - & 1.00\\
GRB $\rm 120212A^{b}$    & 0      & -        & -   &- & -  & -          \\
GRB 120218A  		     & 0     & 0       & -    & - & - & 1.00\\
GRB 120224A              & 0     & 0       & -    &- & - & 1.00\\ [1ex]      
\hline
\end{tabular}}
\tablecomments{\footnotesize  
a. Blank sky was observed over a period of 5 minutes.\\
 b. No blank sky data were recorded due to insufficient storage space at the telescope.} 
\end{table}

\subsection{Discussion}
In this section we  compare our results with the B12 and other previous experiment results, to gain some understanding of the limits on FRB associated with GRBs and to give a quantitative assessment of the conclusions in B12.
     
\subsubsection{Control Observation on Blank Sky}

We have designed our experiment to be  similar to B12 and have also made significant improvements. 
Firstly, unlike B12 we have undertaken control observations using blank patches of sky $\sim$ 2 degrees away from the GRB positions. Secondly, the statistical analysis of the events detected from the blank sky gives  estimates for the number of false positives  due to thermal noise fluctuations and RFI. This method is more reliable when compared to the method adopted by B12 to account for the false positive rate by randomising the spectral channels, using data from the GRB observations. B12 showed that randomising the spectral channels would preserve the properties of thermal noise fluctuations, narrow band and wide band RFI, but the presence of impulsive  RFI that spans multiple channels are destroyed. This will reduce the number of false positive events due to thermal noise fluctuations and RFI.  This effect will be high if the rate of impulsive RFI that spans over multiple channels is high, which leads to an inaccurate estimation of false positives. However, the data from blank sky taken $\sim$ 2 degrees away from the GRB positions preserves all the properties of thermal noise fluctuations and RFI including the periodic and aperiodic RFI that spans multiple channels. Also, the data go through the same RFI excision and detection algorithms as those used for the GRB data sets, which should provide a more reliable estimate of the false positive rates.

\subsubsection{Comparing the experiment parameters with B12}
The average response times for our experiments and B12 are comparable. Table \ref{tbl-5} compares the parameters of the B12 observations and our observations. We calculate that the one sigma flux density for our observations to detect a pulse width of 25 ms is approximately 0.5 Jy and for the B12 observations is 1.07 Jy, indicating that our observations are twice as sensitive as those by B12, (although at different frequencies). The two pulses detected by B12 are of $> 5$ millisecond duration. Therefore, the difference in the minimum time resolution in both experiments is not likely to be significant.

\begin{table}[htp]
\centering  
\caption{\scriptsize \textbf{Observational parameters of this work and B12 \label{tbl-5}}}  
\footnotesize
\begin{tabular}{l c c }  
\hline\hline                       
Parameters & 12m Parkes & 26m Hobart  
\\
\hline\hline 
Central Frequency (MHz)  & 1340 & 2276  \\
Bandwidth (MHz)          & 220  & 64    \\
Time Resolution ($\mu$s) & 64 & 640 \\
$\#$ of Channels           & 600 & 128 \\
$\#$ of Polarisations       & 2    & 2 \\ 
System Equivalent Flux Density-SEFD (Jy) & 3800 & 900 \\
Min detectable flux density-1$\sigma$ (Jy) @ 25ms & 1.07 & 0.5 \\
Beam Full Width at Half Maximum (deg) & 1.28 & 0.35 \\
Average slew time (s) & 175 & 140  \\
S/N threshold & 6 & 5 \\
\hline 
\end{tabular}
\end{table}

\subsubsection{The Similarity of the GRB X-ray light curves}
\begin{figure}[htp]
\center 
\includegraphics[width=4in,height=4.75in]{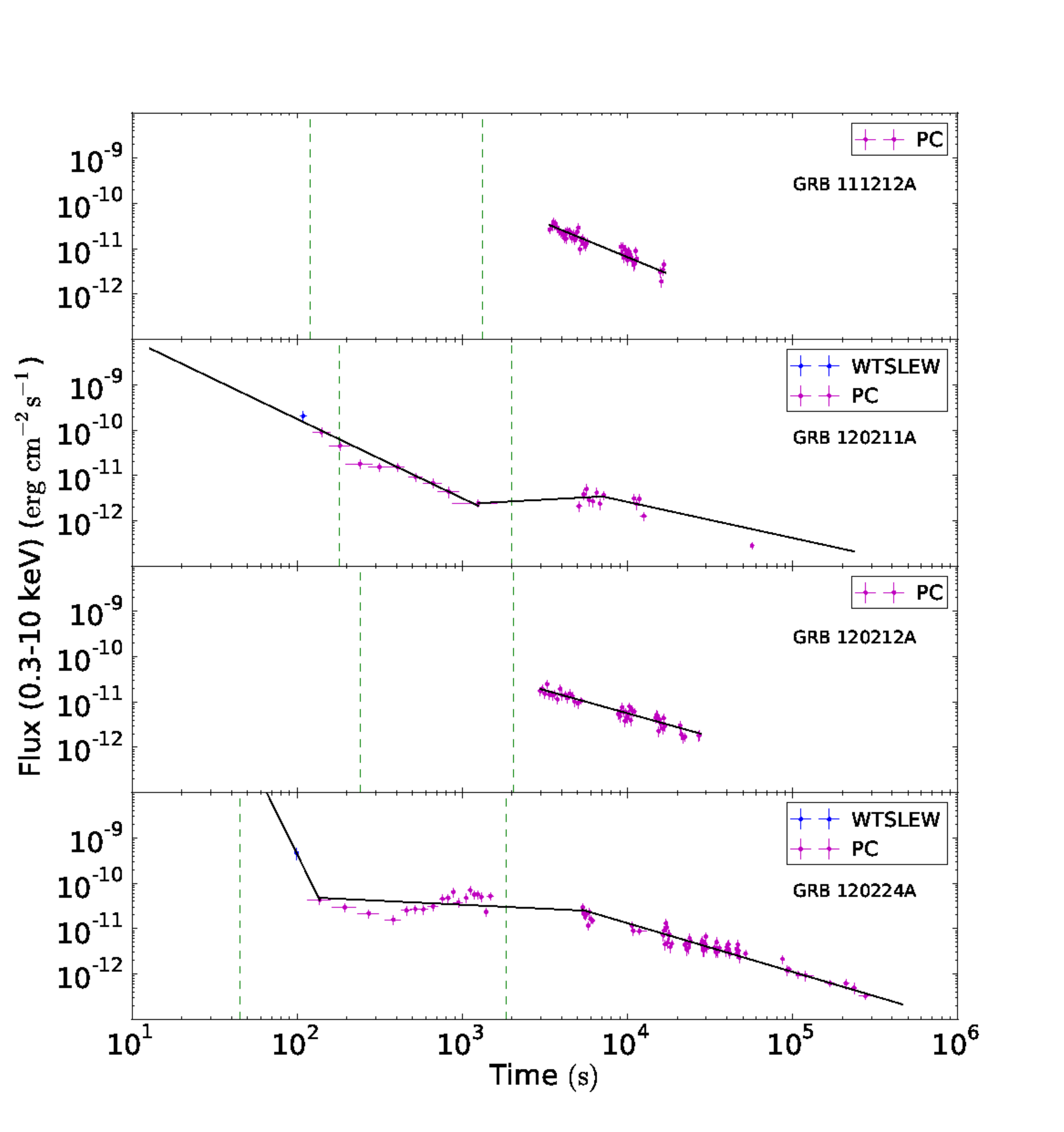}
\caption{\scriptsize The X-ray lightcurves for  the four of the five GRBs observed in our experiment. The observation start ($T_{on}$) and end time are marked as vertical green lines. The black lines are the XRT powerlaw fits. WTSLEW/WT-settling is the window timing mode during the spacecraft slew . PC is the photon counting mode. For GRB 120218A no XRT lightcurve due to sun observing constraints. For GRB 111212A and GRB 120218A no XRT prompt phase information was recorded due to Swift satellite orbital constrains. As the Swift satellite is in a low Earth orbit with the orbit period of 94 minutes, it suffers from 50$\%$ time off target per orbit.  \label{PLightcurve}}
\end{figure}

Figure \ref{PLightcurve} shows the X-ray light curves for the GRBs observed in our experiment and Figure 5 in B12 shows the X-ray light curves for the two GRBs for which radio pulses were detected by B12. The X-ray flux  ($\rm F_{\nu}$, erg $\rm cm^{-2} \, s^{-1} $) is given as $\rm F_{\nu} \propto \nu^{-\beta_{X}} t^{-\alpha_{X}}$ where $\beta_{X}$ is the  spectral index, $\alpha_{X}$ is the temporal decay index, t and $\nu$ are the time and frequency respectively. 

Evaluating Figure \ref{PLightcurve} and Figure 5 in B12, the X-ray telescope (XRT) typically starts observations $\rm \sim 10^{2} \, s$ after the prompt $\gamma$-rays. The X-ray flux  follows a typical pattern which comprises three distinct power-law segments as described by \citet{Nousek2006}, \citet{Zhang2006}, and \citet{O2006}: $\left(1\right)$  initial steep decay slope ($ \rm F_{\nu} \propto t^{-\alpha_{X}}$ with $\rm 3\lesssim \alpha_{X,1} \lesssim 5$ and t $\rm \sim 10^{1}-10^{2}$ s); $\left(2\right)$ a very shallow decay ($\rm 0.5 \lesssim \alpha_{X,2} \lesssim 1.0$ and t $\rm \sim 10^{2}-10^{3}$ s); and  $\left(3\right)$ a steep decay ($ \rm 1.0 \lesssim \alpha_{X,3} \lesssim 1.5$ and t $\rm \sim 10^{3}-10^{4}$ s). These segments are separated by two corresponding break times, $\rm t_{break,1} \lesssim$ 500 s and $\rm 10^{3} \lesssim t_{break,2} \lesssim 10^{4}$ s. 

Table \ref{tbl-6} lists the temporal indices ($\rm \alpha_{X,1}$, $\rm \alpha_{X,2}$,  $\rm \alpha_{X,3}$) and spectral index ($ \rm \beta_{X}$) parameters for all the X-ray light curves in Figure \ref{PLightcurve} and Figure 5 in B12. The temporal indices  and spectral indices  for all the GRBs listed in the Table \ref{tbl-6}  are consistent with the standard GRB X-ray light curve, indicating that the GRBs  are from similar populations (although in many cases we do not see all three power-law segments, due to limited XRT coverage). Therefore we note that the GRBs from which B12 claim FRB emission do not appear different from the GRBs we observed in our experiment. 

\begin{table}[htp]   
\begin{center}  
\caption{\footnotesize Temporal indices ($\rm \alpha_{X,1}$, $\rm \alpha_{X,2}$, $\rm \alpha_{X,3}$) of the X-ray lightcurves for the GRBs observed in our experiment and for two GRBs for which radio pulses were detected B12. $t_{break,1}$ and $t_{break,2}$ are the break times in second on the X-ray lightcurve. $\rm \beta_{X}$ is the spectral index in photon counting (PC) mode \citep{Evans2009} \label{tbl-6}} 
\footnotesize
\begin{tabular}{lcccccc}  
\tableline\tableline
Source & $\rm \alpha_{X,1}$ & $\rm t_{break,1}$ (s) & $\rm \alpha_{X,2}$ & $\rm t_{break,2}$ (s) & $\rm \alpha_{X,3}$ & $\rm \beta_{X}^{\dagger}$ \\ 
\tableline 
GRB $\rm 100704A^{a}$ &3.09 $\pm$ 0.58 & 502 & 0.38 $\pm$ 0.30 & 1157 & 0.90 $\pm$ 0.08 & 1.12 $\pm$ 0.10\\
GRB $\rm 101011A^{a}$ & 2.54 $\pm$ 0.70 & 708 & - & - & 0.89 $\pm$ 0.17 & 0.81 $\pm$ 0.30\\
\hline
GRB $\rm 111212A ^{b}$ & - & - & - & - &1.49 $\pm$ 0.12 & 1.03 $\pm$ 0.14 \\  
GRB 120211A & 1.74 $\pm$ 0.18 & 521 & -0.19 $\pm$ 0.25 & 5082 &  0.79 $\pm$ 0.33 & 0.22 $\pm$ 0.21\\ 
GRB $\rm 120212A^{b}$ & - & - & -&- & 1.01 $\pm$ 0.07 & 1.08 $\pm$ 0.16 \\   
GRB $\rm 120218A^{c}$ & -&-&-&-& -& -\\  
GRB 120224A & 7.38 & 136 & 0.01 $\pm$ 0.10 & 5348 & 1.07 $\pm$ 0.26 & 1.22 $\pm$ 0.14\\
  
\tableline
\tablecomments{\footnotesize a: Two GRBs from B12 experiment for which radio pulses were detected.\\
b : No XRT prompt phase information was recorded due to Swift satellite orbital constrains. As the Swift satellite is in a low Earth orbit with the orbit period of 94 minutes, it suffers from 50$\%$ time off target per orbit. \\
c : No X-ray light curve due to Sun observing constraints. Swift could not slew to the BAT position until 19:14 UT on 20-02-12. \\
$\dagger$ : The information on spectral index $\rm \beta_{X}$ were obtained from the Swift XRT archive.}
\end{tabular}
\end{center}
\end{table}

\begin{figure}[htp]
\center 
\includegraphics[width=4.75in,height=3.0in]{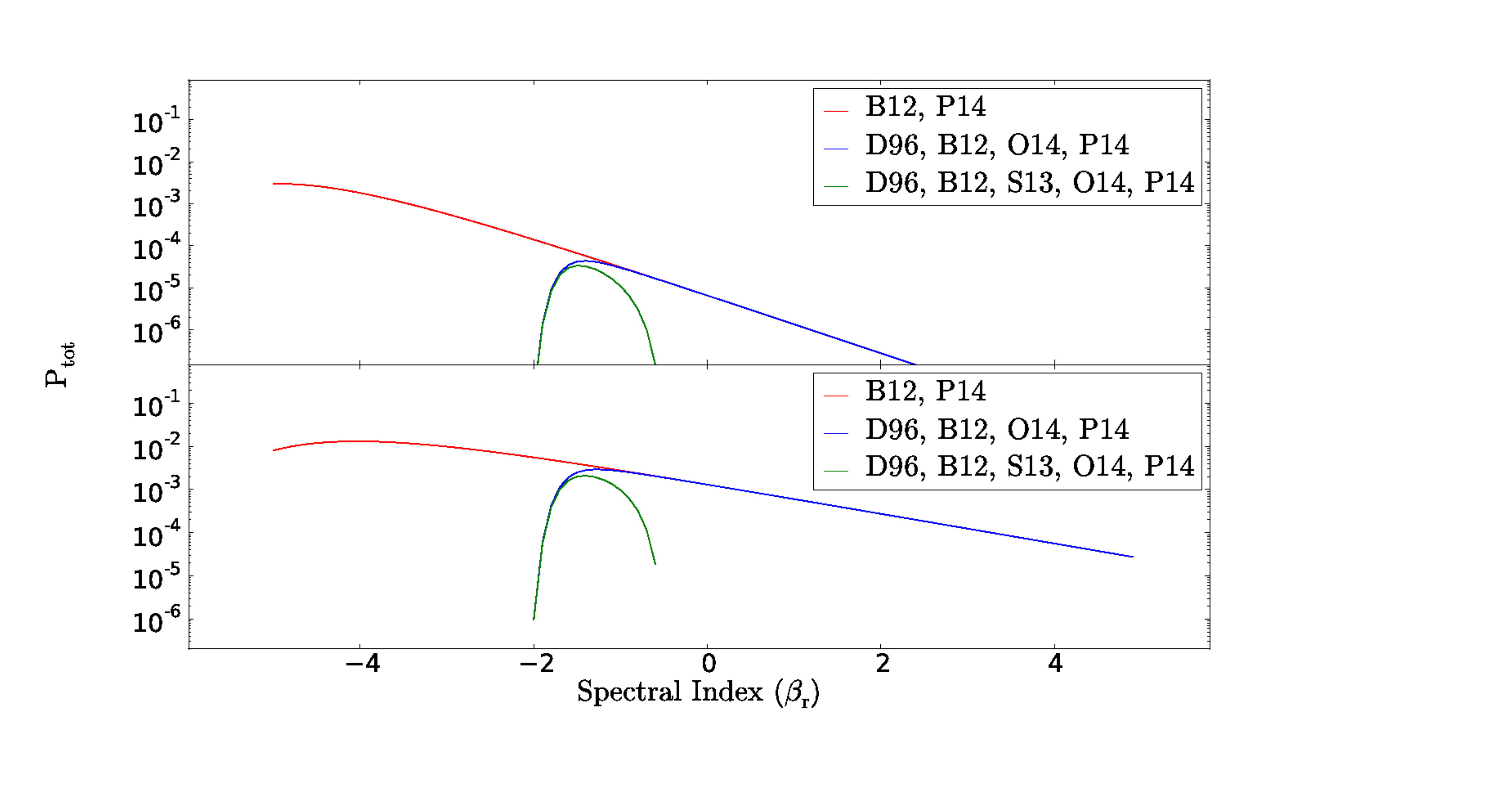}
\caption{\scriptsize \textit{Top panel}: The probability of detecting B12 detecting two events and zero events in other experiments as a function of $\beta_{r}$, spectral index. \textit{Bottom panel}: The probability B12 detecting one real event and zero events in other experiments as a function of $\beta_{r}$, spectral index. $S_{0}$ is the low flux density cutoff. The three curves show the limit imposed by adding more experiments at different frequencies. The red solid line depicts the total probability for P14 and B12 experiments for the range of spectral indices. The blue solid line depicts the total probability of B12 finding events while D96, O14 and P14 experiments have found no events. The green solid line depicts the total probability of  finding events while S13, D96, O14 and P14 experiments have found no events.\label{fig10}}
\end{figure}

\subsubsection{Detection Statistics}

Here we estimate the probability that the combined experiments detected FRB emission, over the combined observations reported in this experiment (hereafter, P14) and in B12. Since many astrophysical objects follow a luminosity function with power-law exponent  \citep{JP2011}, we assume GRBs are distributed homogeneously in a static Euclidean Universe and each GRB generates an FRB with differential number density per unit flux given by,
\begin{equation}
\rm \frac{dN}{dS} \propto S^{-\alpha_{r}} = \mu S^{-\alpha_{r}}.
\end{equation}
where $\rm \alpha_{r}= 2.5$. In order to compare events at different frequencies, we assume the flux density of the FRB has a spectral index $\rm \beta_{r}$ over radio frequencies.

Based on these assumptions, we calculate the total probability of our experiment detecting zero events, and B12 detecting two events out of eight. The total probability, $\rm P_{tot}$, for both experiments is given by the joint probability distribution,                                                                                                                                                                                                                                                                                                                                                                                                                                                                                                                                                                                                                                                                                                                                                                                                                                                                                                                                                                                                                                                                                                                                                                                                                                                                                                                                                                                                                                                                                                                                                                   
\begin{equation}
\rm \rm P_{tot} =  \prod_{i=1}^{2} \rm P(S > S_{min,B12}) \ \times \ \prod_{i=1}^{6} \rm P(S < S_{min,B12}) \ \times \ \prod_{i=1}^{5} \rm P(S < S_{min,P14}),
\end{equation}
where $\rm S$ is the flux density and $\rm S_{min,B12}$ and $\rm S_{min,P14}$ are the threshold flux densities for the B12 experiment and our experiment, respectively.  $\rm P(S > S_{min,B12})$ corresponds to  the probability of detecting events  $\rm > S_{min,B}$  and is given by,
\begin{equation}
\label{eq:prob}
\rm P(S > S_{min,B12}) =  \frac{\int_{S_{min,B12}}^\infty \mu \  S^{-\alpha_{r}} \  dS} {\int_{S_{0}}^\infty \mu \ S^{-\alpha_{r}} \, dS} = \left(\frac{S_{min,B12}}{S_{0}}\right)^{-\alpha_{r} + 1}  ,	
\end{equation}
where the denominator normalises the distribution for a lower limit on the FRB flux density. We chose $\rm S_{0} = 0.189 \, Jy$ to maximise $\rm P_{tot}$ for $\rm \beta_{r}= 0$, hence to maximize the chance that the observed FRBs by B12 are astrophysical. The choice of $\rm S_0$ corresponds to a low flux density cutoff in $\rm dN/dS$ that must exist for the probability of observing an FRB associated with any GRB to not be vanishingly small. Physically this would correspond to a correlation between observed GRB flux density and FRB flux density. Testing the existence of such a correlation is an underlying motivation for this (and similar) experiments.

We note that the observing frequency and S/N threshold are different between our experiment and the B12 experiment (Table \ref{tbl-5}). The intrinsic spectral index is unknown. This difference is accounted for by scaling the flux density with a range of spectral indices $-5 \leq \beta_{r} \leq +5$, given by,
\begin{equation}
\rm S_{min,B12} = 6 \, \sigma_{B12} \left(\frac{\nu_{P14}}{\nu_{B12}}\right)^{\beta_{r}}, 
\end{equation}
where $\rm \sigma_{B12}$ is the minimum detectable flux density at 25 ms in B12 (Table \ref{tbl-5}), $\rm \nu_{P14}$ and $\rm \nu_{B12}$ are the central frequencies in our experiment and B12, respectively.

Next, we estimate $\rm P_{tot}$ in two regimes $\left(1\right)$ the probability of detecting zero events in our experiment, and B12 detecting two events (from eight); $\left( 2\right)$ the probability of detecting zero events in our experiment, and B12 detecting one event (from eight). In the second case we calculate the probability using the equation \ref{eq:prob} just by changing the upper limit on first two terms of equation \ref{eq:prob}. In the former case, the probability $<$ 0.001 (top panel of Figure \ref{fig10}, solid red line) and in the latter case the probability $\rm P_{tot}$ $\lesssim$ 0.01 (bottom panel of Figure \ref{fig10}, solid red line). This indicates that non-detections in our experiment are highly unlikely to be consistent with B12 having obtained two detections.

Next, we now extend the above analysis to include the experiments  of \citet[][hereafter D96]{Dessenne1996}, \citet[][hereafter S13]{Staley2013} and \citet[][hereafter O14]{Obenberger2014}. Table \ref{otherexp} list the parameters for these experiments. The total probability of observed results over all five experiments, is given by,
\begin{equation}
\begin{split}
\rm P_{tot} =  \prod_{i=1}^{2} \rm P(S > S_{min,B12}) \, \times \, \prod_{i=1}^{6} \rm P(S < S_{min,B12}) \, \times \, \prod_{i=1}^{5} \rm P(S < S_{min,P14}) \\
\times \, \prod_{i=1}^{2} \rm P(S < S_{min,D96}) \, \times \, \prod_{i=1}^{34} \rm P(S < S_{min,O14}) \, \times \,\prod_{i=1}^{4} \rm P(S < S_{min,S13}).
\end{split}
\end{equation}

Similar to the above analysis, we calculate the total probability in two regimes: $\left(1 \right)$ the probability of detecting zero events in four different experiments, and B12 detecting two events (from eight); $\left( 2\right)$ the probability of detecting zero events in four different experiments, and B12 detecting one event (from eight). 

To illustrate the effects of adding the low frequency experiments and the high frequency experiments, we combine the experiments in two steps. First, low frequency experiments D96 and O14 are combined with B12 and P14 experiments. Next, both high and low frequency experiments S13, D96 and O14 are combined with B12 and P14 experiments. The blue solid line in the Figure \ref{fig10} depicts the total probability of B12 finding events while D96, O14 and P14 experiments have found no events. The green solid line in the Figure \ref{fig10} depicts the total probability of B12 finding events while S13, D96, O14 and P14 experiment found no events. In both cases the probability of null detection in S13, D96, O14 and P14 experiments, while B12 detecting at least one, is $\rm < 10^{-2}$.

The sensitivities of the S13, D96, O14 experiments are scaled to detect a 25\,ms pulse (Table \ref{otherexp}). In the Figure \ref{fig10}, we note a sharp cut-off in the curve, because at this point the low frequency and high frequency experiments have $\rm S_{min}$ values less than $\rm S_{0}$, which implies that the probability of detecting an event would be vanishingly small. Hence, the low frequency experiments (D96 and O14) would have easily detected a 25\,ms pulse with spectral index $\rm \beta_{r} < -2$ given their detection thresholds, and the high frequency experiment S13 would have easily detected a 25\,ms pulse with the spectral index $\rm \beta_{r} > -0.5$.
 
Finally, we to note that including the other previous low frequency experiments, of \citet{Baird1975}, \citet{Benz1998} and \citet{Balsano1999} in the analysis makes no difference to the results.

\begin{table}[htp]
\caption{\scriptsize Observation parameters of other GRB experiements. \label{otherexp}}  
\centering  
\footnotesize
\begin{tabular}{l l l l c}
\hline\hline
Parameters & D96 & $\rm S13^{\dagger}$ & $\rm O14^{\star}$  \\
\hline\hline 
Central Frequency (MHz)  & 151 & 15270 & 74, 52, 37.9  \\
Bandwidth          &  700 kHz  &   6 GHz  & 75 kHz     \\
Time Resolution (s) & 1.5 & 0.5 & 5    \\
Min detectable flux density-1$\sigma$ (Jy) @ 25ms & 20 & 0.026 & 353, 336, 311 \\
$\#$ of GRBs observed  & 2 &   4  & 12, 5, 17 \\
S/N threshold & 3 & 3 & 5 \\
\hline 
\tablecomments{\footnotesize $\dagger$ The total number of GRBs observed by S13 was 11, out of which only 4 GRBs were observed within 5 min of the GRB notification. Hence we consider only four GRBs in the analysis. \\
$\star$ O14 observed 34 GRBs in total. However the sensitivity of the detector was different for each GRB. Here we have calculated the sensitivity by taking an average over the sensitivities at a given observing frequency.}
\end{tabular}
\end{table}

\section{Conclusion}
\label{sec:conclusion}
This experiment searched for FRB-like emission from GRBs at 2.3 GHz. We observed five GRBs using a 26 m radio telescope, automated to quickly respond to GCN notifications and slew to the source position within minutes. We did not detect any sigma events $> 6\sigma$ similar to B12, which motivated this experiment. Our analysis of events detected at $>$ 5$\sigma$ in our experiment shows that they are consistent with thermal noise fluctuations. Non-detections in our experiment  agree with the lack of consistency between the event rates presented in \citet{Thornton2013} and GRB event rates. 

A joint analysis of our results and four other GRB experiments shows that the B12 results are highly unlikely to be astrophysical. If B12 events are real then the combined analysis constrains the radio spectral index of the events to be $\rm -2 \lesssim \beta \lesssim - 0.5$.

\section{Acknowledgements}

The International Centre for Radio Astronomy Research is a Joint Venture between Curtin
University and The University of Western Australia, funded by the State Government of Western
Australia and the Joint Venture partners. S.J.T is a Western Australian Premiers Research Fellow. R.B.W is supported via the Western Australian Centre of Excellence in Radio Astronomy Science and Engineering. The Centre for All-sky Astrophysics is an Australian Research Council Centre of Excellence, funded by grant CE110001020. D.P is supported by a CSIRS scholarship provided by Curtin University. J. M. is a Super Science Fellow, supported by an ARC grant. 

We would like to thank the referee for his or her useful comments and suggestions for this manuscript. We also thank Peter A. Curran for useful discussions and suggestion on the X-ray light curves and analysis. We would also like to thank Aidan Hotan and Bruce Stansby for providing the Vela data.
This research has made use of the following facilities:
\begin{itemize}
\item The Hobart 26 m radio telescope located at the Mount Pleasant Radio Observatory, operated by the University of Tasmania;
\item The work was supported by iVEC through the use of advanced computing resources located at iVEC@Curtin;
\item The Swift Science Data Centre located at the University of Leicester, UK. 
\end{itemize}

\appendix

\section{Appendix}

\begin{figure}[htp]
\center
\includegraphics[width=4.5in,height=3.0in]{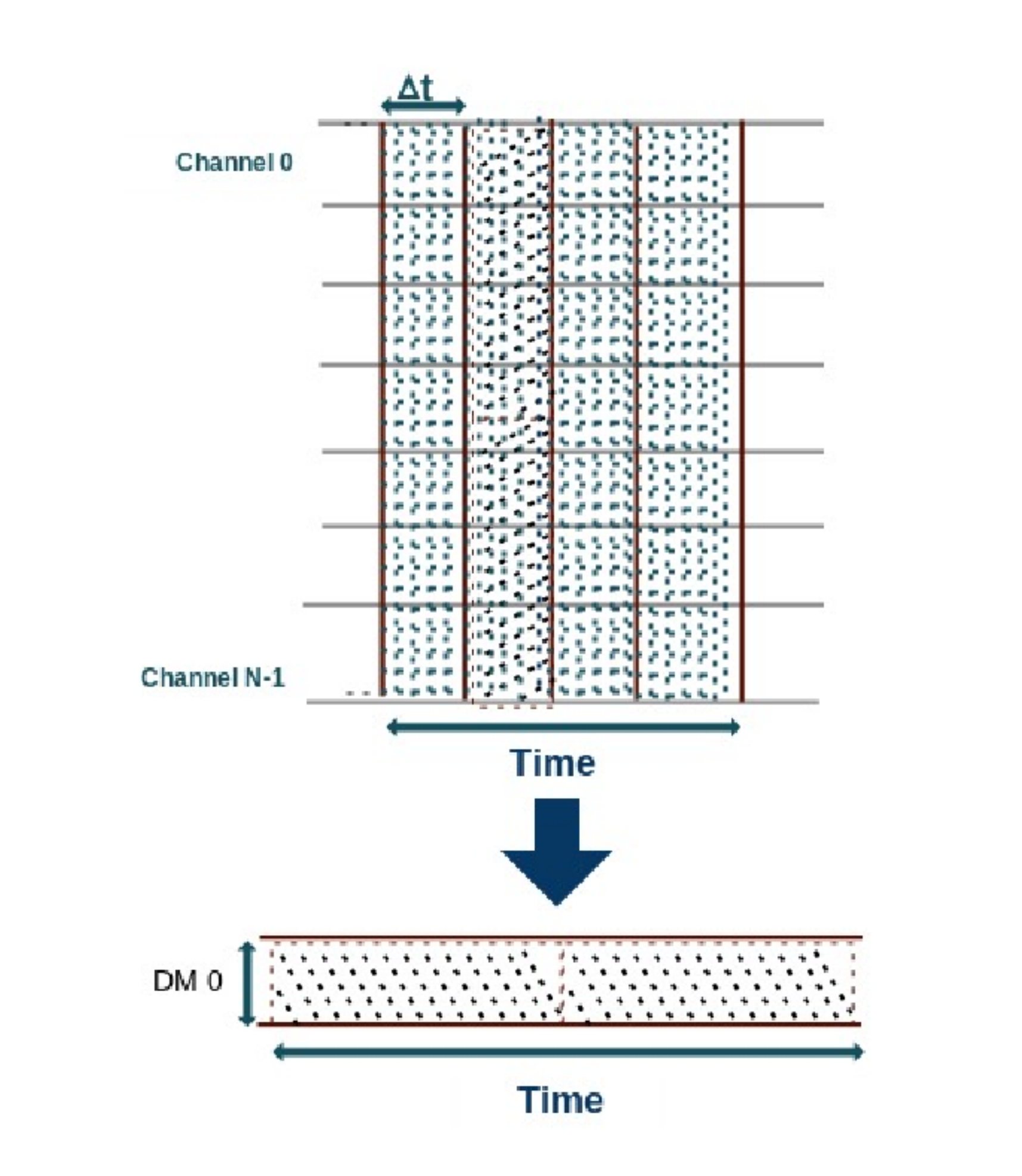}
\caption{\footnotesize \textit{Top panel}: Data incoming into the dedisperser are ordered in frequency and time. The de-dispersion process sums the power samples across spectral channels, with the delay for each channel computed according to the frequency and DM. For DM=0, no delay is applied to the spectral channels and all the power samples are summed to form a time series (Bottom panel). \label{DM0}}
\end{figure}

Data incoming into the dedisperser are discrete samples in frequency and time. The de-dispersion process sums the power samples across spectral channels, with the delay for each channel computed according to the frequency and DM (Figure \ref{DM0}). Under normal circumstances the power level in each spectral channel follows a Gaussian distribution.

For low DMs, there is a $\sim$50$\%$ overlap between the power samples that contribute to consecutive DM trials, at the same time. Here we demonstrate that this overlap does not correspond to a high level of correlation in detections between neighbouring DM steps for a large (e.g $\rm 5 \sigma$) detection threshold, and therefore that each DM step can be considered independent.

Let us consider a DM trial where the power samples in the channels are normally distributed across the spectrum. The variance of summing $N_{\rm chan}$ independent samples equals $N_{\rm chan} \, \sigma_{i}^{2}$ and the standard deviation, $\sigma{=}\rm \sqrt{N_{\rm chan}} \, \sigma_{i}$, where $\sigma_{i}$ is the noise level in a given power sample. 

The probability of obtaining an event above a threshold, $X$, due to random addition of power sample fluctuations is given by the CDF;
\begin{equation}
\rm P \rm (I > X\sigma) = \frac{1}{2} \, - \, \frac{1}{2} \,  \rm erf\left[\frac{X \, \sigma}{\sqrt{2} \, \sigma}\right],
\label{gaussian_cdf}
\end{equation}
Substituting $\rm \sigma = \sqrt{N_{\rm chan}} \, \sigma_{i}$ in Equation \ref{gaussian_cdf}, gives the CDF in terms of the noise level in each power sample;
\begin{equation}
\rm P \rm (I > X\sigma) = \frac{1}{2} \, - \, \frac{1}{2} \,  \rm erf\left[\frac{X \, \sqrt{N_{\rm chan}} \: \sigma_{i}}{\sqrt{2} \: \sqrt{N_{\rm chan}} \: \sigma_{i}}\right].
\label{gaussian_cdf_sigma}
\end{equation}
The probability that we obtain a detection above our experimental threshold (5$\sigma$) from an independent DM step is given by,
\begin{equation}
\rm \left[ P(I > 5\sigma)\right] = \frac{1}{2} \, - \, \frac{1}{2} \,  \rm erf\left[\frac{5 \: \sqrt{128} \: \sigma_{i}}{\sqrt{2} \: \sqrt{128} \: \sigma_{i}}\right] = 2.87 \times 10^{-7}.
\label{gaussian_cdf_value}
\end{equation}
where $N_{\rm chan}=128$.

\begin{figure}[htp]
\center
\includegraphics[width=3.5in,height=2.4in]{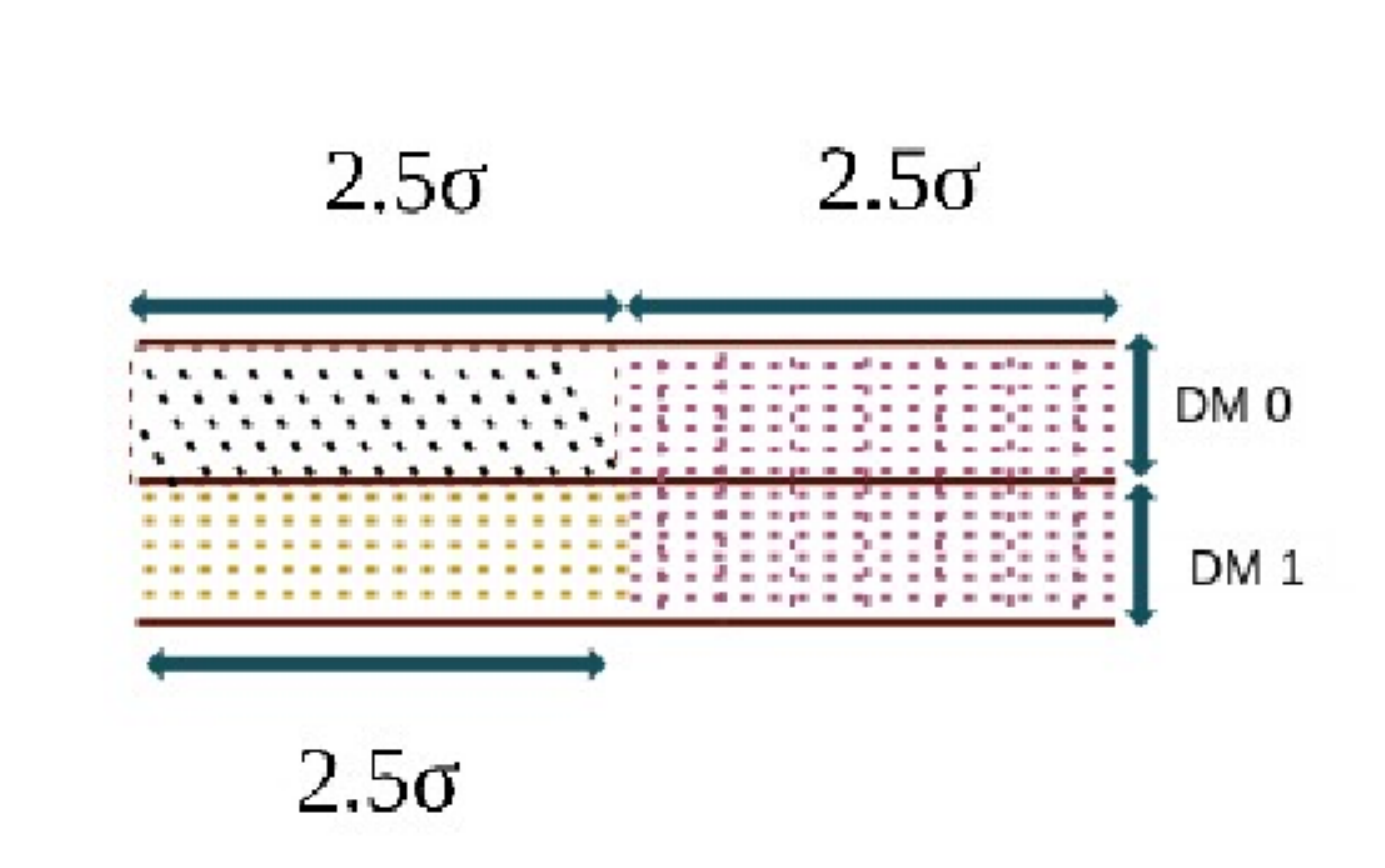}
\caption{\footnotesize Figure illustrates that $\sim 50\%$ of power samples at DM of 1 overlaps with DM of 0 represented in dotted pink lines. And rest are the non-overlapping samples in the DM of 0 and 1 repersented in dotted black and yellow lines.\label{two_DM}}
\end{figure}

For low DMs, we note that due to our choice of progression in the DM steps, there is a $\sim$50$\%$ overlap between the power samples that contribute to consecutive DM trials, at the same time.  When there are shared samples between consecutive DM steps, those samples contribute the same power to both DM trials. There are many permutations for how power can be distributed in each set of power samples to combine to yield two detections above $\rm X\sigma$. It can be shown that the power distribution with the highest joint probability over the two DM trials corresponds to the case shown in Figure \ref{two_DM}, where the shared power samples contribute proportionally to the power required to meet the threshold. Under this scenario, \textit{each} of the three sets of $\rm N_{\rm overlap}=64$ samples contributes 2.5$\rm \sigma=$2.5$\rm \sqrt{N_{\rm chan}}\sigma_{i}$. The probability that each set obtains a detection $\rm > 2.5\sigma$ is given by,
\begin{equation}
\rm P_{\rm set}\rm (I > X_{\rm set}\sigma) = \frac{1}{2} \, - \, \frac{1}{2} \,  \rm erf\left[\frac{X_{\rm set} \, \sqrt{N_{\rm chan}} \, \sigma}{\sqrt{2 \: 	N_{\rm overlap}} \,\sigma}\right],
\end{equation}
with $X_{\rm set}=X N_{\rm overlap}/N_{\rm chan}=2.5$, giving,
\begin{equation}
\rm P \rm (I > X\sigma) = \frac{1}{2} \, - \, \frac{1}{2} \,  \rm erf\left[\frac{X \, \sqrt{N_{\rm overlap}}}{\sqrt{2 \: 	N_{\rm chan}}}\right].
\end{equation}

When 50$\%$ of samples overlap and the threshold, $\rm X=5$, the joint probability of obtaining detections in two consecutive DM trials is given by the product of the probability for each of the three sets;
\begin{equation}
\rm P(I > 5\sigma) = \left[P_{\rm set}(I > 2.5\sigma \, ; \, N_{overlap} = 64)\right]^{3}  = 1.7 \times 10^{-9}.
\end{equation}
This is $\sim$100 times smaller than the probability of detecting a single event (Equations \ref{gaussian_cdf}-\ref{gaussian_cdf_value}). Therefore, although the power samples contributing to the consecutive DM trials are 50$\%$ correlated, due to the improbability of obtaining a single 5$\sigma$ detection, the joint probability of obtaining two detections is very small. We therefore conclude that the DM trials can be considered statistically independent.

\end{document}